\documentclass[12pt,english]{article}
\usepackage[T1]{fontenc}
\usepackage[latin9]{inputenc}
\usepackage{geometry}
\usepackage{array}
\usepackage{float}
\usepackage{bm}
\usepackage{multirow}
\usepackage{amsmath}
\usepackage{amsthm}
\usepackage{graphicx}
\usepackage{rotfloat}
\usepackage[authoryear]{natbib}

\makeatletter

\providecommand{\tabularnewline}{\\}

\numberwithin{equation}{section}

\@ifundefined{date}{}{\date{}}
\makeatother

\usepackage{babel}
\begin{document}
\title{Bayes factors for accelerated life testing models}
\author{Neill Smit \thanks{{\small{}School of Mathematical and Statistical Sciences, North-West
University, Vanderbijlpark, South Africa}, neillsmit1@gmail.com} $\;$ and Lizanne Raubenheimer \thanks{{\small{}Department of Statistics, Rhodes University, Makhanda, South
Africa}}}
\maketitle
\begin{abstract}
In Bayesian accelerated life testing, the most used tool for model
comparison is the deviance information criterion. An alternative and
more formal approach is to use Bayes factors to compare models. However,
Bayesian accelerated life testing models with more than one stressor
often have mathematically intractable posterior distributions and
Markov chain Monte Carlo methods are employed to obtain posterior
samples to base inference on. The computation of the marginal likelihood
is challenging when working with such complex models. In this paper,
methods for approximating the marginal likelihood and the application
thereof in the accelerated life testing paradigm are explored for
dual-stress models.

\end{abstract}

\section{Introduction\label{sec:BF_INTRODUCTION}}

Bayesian model selection is an important aspect of any Bayesian analysis
and aids the researcher in determining which of the proposed models
should be used. Various approaches towards Bayesian model selection
are discussed in Kadane \& Lazar$^{1}$. The different perspectives
to model selection discussed include Bayes factors, Bayesian model
averaging, methods used for Bayesian linear models, and predictive
methods. The authors also state that model selection tools should
rather be used to identify inappropriate models, instead of attempting
so select a single best model.

Bayesian accelerated life testing (ALT) models with more than one
stressor often have mathematically intractable posterior distributions
and Markov chain Monte Carlo (MCMC) methods are employed to obtain
posterior samples to base inference on. For this reason, and the fact
that it is a standard output in many Bayesian data analysis software
packages, the deviance information criterion (DIC) is a very popular
tool for model comparison. It can be argued that the more formal and
traditional approach for Bayesian model selection would be the use
of Bayes factors (Spiegelhalter \textit{et al}.$^{2}$, Upadhyay \&
Mukherjee$^{3}$).

The authors in Spiegelhalter \textit{et al}.$^{2}$ explain that the
DIC is intended as an alternative to Bayes factors and that there
are situations where the DIC may be a more appropriate tool for model
comparison. Bayes factors are the most suited when the complete set
of possible models can be specified and the true model is included
in this set, whereas the DIC does not have this requirement (Bernardo
\& Smith$^{4}$, Spiegelhalter \textit{et al}.$^{2}$). In a discussion
on the DIC, the authors state that Bayes factors and the DIC are intended
for different purposes, and can thus lead to different conclusions
regarding model selection. Bayes factors consider how well the prior
predicts the observed data, whereas the DIC considers how well the
posterior will predict future data by means of the same mechanism
that resulted in the observed data (Spiegelhalter \textit{et al}.$^{2}$).

In this paper, model selection in the Bayesian ALT setup is considered
by comparing a generalized Eyring-Weibull (GEW) model and a generalized
Eyring-Birnbaum-Saunders (GEBS) model via the DIC and Bayes factors.
Due to the mathematically intractable posterior distributions of these
models, the computation of the marginal likelihood, needed for the
calculation of the Bayes factor, is also complicated. Methods for
approximating the marginal likelihood, without further complicating
the MCMC sampling process, are discussed. The methods considered include
a simple Monte Carlo estimator, the harmonic mean estimator, the Laplace-Metropolis
estimator, and a posterior predictive density (PPD) estimate for posterior
Bayes factors.

\section{Bayesian Model Selection\label{sec:BF_MODEL_SELECTION}}

\subsection{Deviance Information Criterion}

The DIC, proposed by Spiegelhalter \textit{et al}.$^{2}$, is a widely
used measure for Bayesian model comparison. It is used to assess both
the goodness-of-fit and the complexity of the model. The model with
a significantly lower DIC value will usually be the preferred model
to use, but there are other considerations as well. The authors state
that there is no specific rule of thumb on what constitutes a significant
difference in DIC, but that guidelines proposed by Burnham \& Anderson$^{5}$
on the Akaike information criterion (AIC) also seem to work well for
the DIC.

The formulation of the DIC follows on the work of Akaike$^{6}$ and
is based on using the posterior mean deviance as a measure of fit,
and a new complexity measure called the \textit{effective number of
parameters}. For a parameter vector $\bm{\theta}$, with likelihood
function $L\left(\text{\ensuremath{\underline{x}}}\left|\bm{\theta}\right.\right)$,
the deviance can be defined as
\[
D\left(\bm{\theta}\right)=-2\ln\left[L\left(\text{\ensuremath{\underline{x}}}\left|\bm{\theta}\right.\right)\right].
\]
Denote the posterior mean of the deviance by $\overline{D}$, and
let
\[
\hat{D}\left(\bm{\overline{\theta}}\right)=-2\ln\left[L\left(\text{\ensuremath{\underline{x}}}\left|\bm{\overline{\theta}}\right.\right)\right],
\]
be a point estimate for the deviance, with $\bm{\overline{\theta}}$
being the posterior mean of $\bm{\theta}$. The DIC can then be calculated
as
\[
\textrm{DIC}=\overline{D}+p_{D},
\]
where $p_{D}$ is the effective number of parameters given by $p_{D}=\overline{D}-\hat{D}\left(\bm{\overline{\theta}}\right)$.
It is shown in Spiegelhalter \textit{et al}.$^{2}$ that the DIC is
approximately equivalent to the AIC for models with very weak prior
information.

The DIC is a very popular choice for model comparison in the Bayesian
ALT setup, particularly when working with complicated models where
MCMC methods are used to obtain posterior samples for inference (see,
for example, Barriga \textit{et al}.$^{7}$, Soyer \textit{et al}$^{8}$,
and Upadhyay \& Mukherjee$^{3}$). The reason for this is that the
DIC can easily be obtained from the MCMC output (see Spiegelhalter
\textit{et al}.$^{2}$), and it is also provided as a standard output
in some well-known Bayesian data analysis software such as WinBUGS/OpenBUGS
and JAGS.

\subsection{Bayes Factors}

The foundation for Bayesian hypothesis testing via Bayes factors was
developed by Jeffreys$^{9}$. The author referred to his methods as
``significance tests'' and presented them as an alternative to $p$-values.
The approach compares predictions made by two competing scientific
theories by defining statistical models for each theory and calculating
the posterior probability that one of the theories is correct (Kass
\& Raftery$^{10}$).

Denote by $f\left(m_{i}\left|\underline{x}\right.\right)$ and $f\left(m_{j}\left|\underline{x}\right.\right)$
the posterior model probabilities for models $m_{i}$ and $m_{j}$,
respectively. The comparison of two models $m_{i}$ and $m_{j}$,
in the Bayesian setup, is conducted by means of the posterior odds
for model $m_{i}$ against model $m_{j}$. This is given by
\[
PO_{ij}=\frac{f\left(m_{i}\left|\underline{x}\right.\right)}{f\left(m_{j}\left|\underline{x}\right.\right)}=\frac{f\left(\underline{x}\left|m_{i}\right.\right)}{f\left(\underline{x}\left|m_{j}\right.\right)}\times\frac{f\left(m_{i}\right)}{f\left(m_{j}\right)}=B_{ij}\times\frac{f\left(m_{i}\right)}{f\left(m_{j}\right)},
\]
where $f\left(\underline{x}\left|m_{g}\right.\right)$, $g=i,j$,
are the marginal likelihoods and $f\left(m_{g}\right)$, $g=i,j,$
are the prior model probabilities of the two models. The above expression
is often summarized as
\[
{\rm Posterior\ odds}={\rm Bayes\ factor}\times{\rm Prior\ odds}.
\]
In most situations no prior information will be available regarding
the model structure, in which case the prior model probabilities are
set equal (see, for example, Ntzoufras$^{11}$). This results in using
the Bayes factor for hypothesis testing, which is also extended to
model selection. It is important to note that Bayes factors can be
used to evaluate evidence \textit{against} the null hypothesis or
\textit{in favour} of the null hypothesis, which is not possible in
classical hypothesis testing.

The Bayes factor $B_{ij}$ of model $m_{i}$ versus model $m_{j}$
is defined as the ratio of the marginal likelihoods $f\left(\underline{x}\left|m_{i}\right.\right)$
and $f\left(\underline{x}\left|m_{j}\right.\right)$. This can be
written as
\begin{equation}
B_{ij}=\frac{f\left(\underline{x}\left|m_{i}\right.\right)}{f\left(\underline{x}\left|m_{j}\right.\right)}=\frac{\int f\left(\underline{x}\left|\bm{\theta_{m_{i}}},m_{i}\right.\right)f\left(\bm{\theta_{m_{i}}}\left|m_{i}\right.\right)d\bm{\theta_{m_{i}}}}{\int f\left(\underline{x}\left|\bm{\theta_{m_{j}}},m_{j}\right.\right)f\left(\bm{\theta_{m_{j}}}\left|m_{j}\right.\right)d\bm{\theta_{m_{j}}}},\label{eq:BF_FORMULA}
\end{equation}
where $f\left(\underline{x}\left|\bm{\theta_{m_{g}}},m_{g}\right.\right)$,
$g=i,j$, is the likelihood of model $m_{g}$ with parameter vector
$\bm{\theta_{m_{g}}}$, and $f\left(\bm{\theta_{m_{g}}}\left|m_{g}\right.\right)$,
$g=i,j$, is the prior imposed on $\bm{\theta_{m_{g}}}$ under model
$m_{g}$.

Interpreting Bayes factors in half-units on the $\log_{10}$-scale
is suggested in Jeffreys$^{9}$, where the author provides a table
for interpreting Bayes factor values. Kass \& Raftery$^{10}$ suggest
a modified version of these interpretations by rather considering
twice the natural logarithm of the Bayes factor. By doing this, Bayes
factors are interpreted on the same scale as the likelihood ratio
test statistic. The interpretation of Bayes factors suggested in Kass
\& Raftery$^{10}$ are used in this paper.

In some cases, the marginal likelihoods in (\ref{eq:BF_FORMULA})
can be computed analytically (see, for example, DeGroot$^{12}$, Zellner$^{13}$).
More often than not, the marginal likelihood must be estimated. Ntzoufras$^{11}$
explains that numerous other versions of Bayes factors as well as
alternative model selection approaches have also been developed. Other
popular types of Bayes factors include pseudo Bayes factors, resulting
from the work of Geisser \& Eddy$^{14}$, posterior Bayes factors,
presented in Aitkin$^{15}$, fractional Bayes factors, introduced
in O'Hagan$^{16}$, and intrinsic Bayes factors, presented in Berger
\& Pericchi$^{17}$.

Suppose we compare two models, $m_{i}$ and $m_{j}$, where the prior
model probabilities are denoted by $f\left(m_{g}\right)$, $g=i,j.$
From Bayes' theorem we then have that the posterior model probability
for model $m_{i}$ can be written as
\begin{eqnarray*}
f\left(m_{i}\left|\underline{x}\right.\right) & = & \frac{f\left(\underline{x}\left|m_{i}\right.\right)f\left(m_{i}\right)}{f\left(\underline{x}\left|m_{i}\right.\right)f\left(m_{i}\right)+f\left(\underline{x}\left|m_{j}\right.\right)f\left(m_{j}\right)}\\
 & = & \frac{f\left(m_{i}\right)\int f\left(\underline{x}\left|\bm{\theta_{m_{i}}},m_{i}\right.\right)f\left(\bm{\theta_{m_{i}}}\left|m_{i}\right.\right)d\bm{\theta_{m_{i}}}}{f\left(m_{i}\right)\int f\left(\underline{x}\left|\bm{\theta_{m_{i}}},m_{i}\right.\right)f\left(\bm{\theta_{m_{i}}}\left|m_{i}\right.\right)d\bm{\theta_{m_{i}}}+\left[1-f\left(m_{i}\right)\right]\int f\left(\underline{x}\left|\bm{\theta_{m_{j}}},m_{j}\right.\right)f\left(\bm{\theta_{m_{j}}}\left|m_{j}\right.\right)d\bm{\theta_{m_{j}}}},
\end{eqnarray*}
where $f\left(\underline{x}\left|m_{g}\right.\right)$, $g=i,j$,
is the marginal likelihoods, $f\left(\underline{x}\left|\bm{\theta_{m_{g}}},m_{g}\right.\right)$,
$g=i,j$, denotes the likelihood of model $m_{g}$ with parameter
vector $\bm{\theta_{m_{g}}}$, and $f\left(\bm{\theta_{m_{g}}}\left|m_{g}\right.\right)$,
$g=i,j$, is the prior imposed on $\bm{\theta_{m_{g}}}$ under model
$m_{g}$. The posterior model probability for model $m_{i}$ can easily
be computed by re-writing it in terms of Bayes factors as
\begin{eqnarray*}
f\left(m_{i}\left|\underline{x}\right.\right) & = & \left[1+\frac{1-f\left(m_{i}\right)}{f\left(m_{i}\right)}B_{ij}^{-1}\right]^{-1},
\end{eqnarray*}
where $B_{ij}$ is the Bayes factor of model $m_{i}$ versus model
$m_{j}$.

\section{Generalized Eyring ALT Models\label{sec:BF_GE_MODELS}}

\noindent In ALT, items are tested in an environment that is more
severe than their normal operating environment, in order to induce
early failures. This is performed by applying accelerated levels of
stressors such as temperature, voltage or humidity to the items. The
life characteristics under normal operating conditions can then be
extrapolated from these accelerated failure times, by means of a functional
relationship known as a time transformation function.

Consider two stressors, one thermal and one non-thermal. Indicate
the $k$ distinct accelerated levels of the stressors by $\left\{ T_{i},S_{i}\right\} ,i=1,\ldots,k$,
where $T_{i},i=1,\ldots,k$, is the accelerated levels of the thermal
stressor and $S_{i},i=1,\ldots,k$ is the accelerated levels of the
non-thermal stressor. An item is exposed to the constant application
of a specific stress level combination $\left\{ T_{i},S_{i}\right\} $.

Suppose that $n_{i}$ items are tested at each of the $k$ different
stress levels and the test is truncated at time $\tau_{i}$. Denote
the failure times by $x_{ij},j=1,\ldots,n_{i},i=1,\ldots,k.$ The
likelihood function, in general, is then given by
\[
L\left(\underline{x}\left|\theta_{1},\theta_{2},\theta_{3},\theta_{4},\beta\right.\right)=\prod_{i=1}^{k}\left[\prod_{j=1}^{r_{i}}f\left(x_{ij}\left|\theta_{1},\theta_{2},\theta_{3},\theta_{4},\beta\right.\right)\right]\left[R\left(\tau_{i}\right)\right]^{n_{i}-r_{i}},
\]
where $R(\cdot)$ is the reliability function. Note that for complete
samples $r_{i}=n_{i}$. For type-I censoring $r_{i}$ is the number
of failures that occur before censoring time $\tau_{i}$, where $\tau_{i}<\infty,i=1,\ldots,k$
is predetermined censoring times for the $k$ different stress levels.
For type-II censoring $\tau_{i}=x_{i(r_{i})}$, where $x_{i(r_{i})}$
is the $r_{i}^{th}$ ordered failure time, and $r_{i},i=1,\ldots,k$
is the pre-chosen number of failures after which censoring occurs
for the $k$ different stress levels. Smit$^{18}$ and Smit \& Raubenheimer$^{19}$
investigates Bayesian Eyring ALT models using the Birnbaum-Saunders
and Weibull distributions as the lifetime models.

\subsection{The GEBS model}

Let $X$ be a continuous random variable that follows a Birnbaum-Saunders
distribution with shape parameter $\alpha$ and scale parameter $\beta$
$\left(\alpha>0,\beta>0\right)$. The PDF of $X$ is then given by

\begin{equation}
f\left(x\left|\alpha,\beta\right.\right)=\frac{x+\beta}{2\sqrt{2\pi}\alpha\sqrt{\beta}\sqrt{x_{i}^{3}}}\exp\left[-\frac{1}{2\alpha^{2}}\left(\frac{x}{\beta}+\frac{\beta}{x}-2\right)\right],\quad x>0,\label{eq:BS_PDF}
\end{equation}
and the reliability function by
\[
R\left(\tau\right)=1-\Phi\left[\frac{1}{\alpha}\left(\sqrt{\frac{\tau}{\beta}}-\sqrt{\frac{\beta}{\tau}}\right)\right],
\]
where $\Phi\left(\cdot\right)$ is the CDF of the standard normal
distribution. A common practice in ALT is to assume that the Birnbaum-Saunders'
scale parameter $\beta$ depends on the stress levels, whereas the
shape parameter $\alpha$ does not (see, for example, Owen \& Padgett$^{20}$,
Upadhyay \& Mukherjee$^{3}$, Sun \& Shi$^{21}$, Sha$^{22}$). This
assumption relates to the failure mechanism remaining constant for
all stress levels. The reparameterization of $\beta$ given by the
generalized Eyring model is
\begin{equation}
\beta_{i}=\frac{1}{T_{i}}\exp\left(\theta_{1}+\frac{\theta_{2}}{T_{i}}+\theta_{3}V_{i}+\frac{\theta_{4}V_{i}}{T_{i}}\right),\label{eq:BS_EYRING}
\end{equation}
where $\theta_{1},$ $\theta_{2},$ $\theta_{3},$ and $\theta_{4}$
are unknown parameters, and $V_{i}$ is a function of the non-thermal
stressor $S_{i}$ (Escobar \& Meeker$^{23}$). From (\ref{eq:BS_PDF})
and (\ref{eq:BS_EYRING}) it follows that the Birnbaum-Saunders PDF
of a lifetime subjected to the $i^{th}$ stress level, can be written
as
\begin{eqnarray}
f\left(x_{i}\left|\alpha,\theta_{1},\theta_{2},\theta_{3},\theta_{4}\right.\right) & = & \frac{x_{i}+\frac{1}{T_{i}}\exp\left(\theta_{1}+\frac{\theta_{2}}{T_{i}}+\theta_{3}V_{i}+\frac{\theta_{4}V_{i}}{T_{i}}\right)}{2\sqrt{2\pi}\alpha\sqrt{x_{i}^{3}}\sqrt{\frac{1}{T_{i}}\exp\left(\theta_{1}+\frac{\theta_{2}}{T_{i}}+\theta_{3}V_{i}+\frac{\theta_{4}V_{i}}{T_{i}}\right)}}\nonumber \\
 &  & \times\exp\left\{ -\frac{1}{2\alpha^{2}}\left[x_{i}T_{i}\exp\left(-\theta_{1}-\frac{\theta_{2}}{T_{i}}-\theta_{3}V_{i}-\frac{\theta_{4}V_{i}}{T_{i}}\right)\right.\right.\nonumber \\
 &  & \quad\left.\left.+\frac{1}{x_{i}T_{i}}\exp\left(\theta_{1}+\frac{\theta_{2}}{T_{i}}+\theta_{3}V_{i}+\frac{\theta_{4}V_{i}}{T_{i}}\right)-2\right]\right\} .\label{eq:GEBS_PDF}
\end{eqnarray}
The corresponding reliability function at some time $\tau$ is given
by
\begin{eqnarray}
R(\tau) & = & 1-\Phi\left\{ \frac{1}{\alpha}\left[\sqrt{\tau T_{i}\exp\left(-\theta_{1}-\frac{\theta_{2}}{T_{i}}-\theta_{3}V_{i}-\frac{\theta_{4}V_{i}}{T_{i}}\right)}\right.\right.\nonumber \\
 &  & \qquad\left.\left.-\sqrt{\frac{1}{\tau T_{i}}\exp\left(\theta_{1}+\frac{\theta_{2}}{T_{i}}+\theta_{3}V_{i}+\frac{\theta_{4}V_{i}}{T_{i}}\right)}\right]\right\} .\label{eq:GEBS_RF}
\end{eqnarray}
From (\ref{eq:GEBS_PDF}) and (\ref{eq:GEBS_RF}), it follows that
the likelihood function for the GEBS model can be written as
\begin{eqnarray}
L\left(\underline{x}\left|\alpha,\theta_{1},\theta_{2},\theta_{3},\theta_{4}\right.\right) & = & \left(2\sqrt{2\pi}\alpha\right)^{-\sum_{i=1}^{k}r_{i}}\exp\left(\frac{1}{\alpha^{2}}\sum_{i=1}^{k}r_{i}\right)\nonumber \\
 &  & \times\left[\prod_{i=1}^{k}\prod_{j=1}^{r_{i}}\frac{x_{ij}+\frac{1}{T_{i}}\exp\left(\theta_{1}+\frac{\theta_{2}}{T_{i}}+\theta_{3}V_{i}+\frac{\theta_{4}V_{i}}{T_{i}}\right)}{x_{ij}^{\frac{3}{2}}}\right]\nonumber \\
 &  & \times\exp\left\{ -\frac{1}{2\alpha^{2}}\sum_{i=1}^{k}\left[T_{i}\exp\left(-\theta_{1}-\frac{\theta_{2}}{T_{i}}-\theta_{3}V_{i}-\frac{\theta_{4}V_{i}}{T_{i}}\right)\sum_{j=1}^{r_{i}}x_{ij}\right]\right.\nonumber \\
 &  & \left.\quad-\frac{1}{2\alpha^{2}}\sum_{i=1}^{k}\left[\frac{1}{T_{i}}\exp\left(\theta_{1}+\frac{\theta_{2}}{T_{i}}+\theta_{3}V_{i}+\frac{\theta_{4}V_{i}}{T_{i}}\right)\sum_{j=1}^{r_{i}}\frac{1}{x_{ij}}\right]\right\} \nonumber \\
 &  & \times\prod_{i=1}^{k}\left\{ \left[T_{i}\exp\left(-\theta_{1}-\frac{\theta_{2}}{T_{i}}-\theta_{3}V_{i}-\frac{\theta_{4}V_{i}}{T_{i}}\right)\right]^{\frac{r_{i}}{2}}\right.\nonumber \\
 &  & \quad\times\left[1-\Phi\left(\frac{1}{\alpha}\left(\sqrt{\tau_{i}T_{i}\exp\left(-\theta_{1}-\frac{\theta_{2}}{T_{i}}-\theta_{3}V_{i}-\frac{\theta_{4}V_{i}}{T_{i}}\right)}\right.\right.\right.\nonumber \\
 &  & \left.\left.\left.\left.\quad\quad-\sqrt{\frac{1}{\tau_{i}T_{i}}\exp\left(\theta_{1}+\frac{\theta_{2}}{T_{i}}+\theta_{3}V_{i}+\frac{\theta_{4}V_{i}}{T_{i}}\right)}\right)\right)\right]^{n_{i}-r_{i}}\right\} .\label{eq:GEBS_LIKELIHOOD}
\end{eqnarray}
Assume independent priors on the model parameters $\theta_{1},$ $\theta_{2},$
$\theta_{3},$ $\theta_{4}$ and $\alpha$ (see, for example, Upadhyay
\& Mukherjee$^{3}$), which leads to the joint prior distribution
being given by
\begin{eqnarray}
\pi\left(\alpha,\theta_{1},\theta_{2},\theta_{3},\theta_{4}\right) & = & \pi\left(\alpha\right)\pi\left(\theta_{1}\right)\pi\left(\theta_{2}\right)\pi\left(\theta_{3}\right)\pi\left(\theta_{4}\right).\label{eq:GENERAL_PRIOR}
\end{eqnarray}
The joint posterior distribution is then given by
\[
\pi\left(\alpha,\theta_{1},\theta_{2},\theta_{3},\theta_{4}\left|\underline{x}\right.\right)\propto L\left(\underline{x}\left|\alpha,\theta_{1},\theta_{2},\theta_{3},\theta_{4}\right.\right)\pi\left(\alpha,\theta_{1},\theta_{2},\theta_{3},\theta_{4}\right).
\]
For the GEBS model, gamma priors are imposed on all the parameters,
therefore
\begin{eqnarray*}
\alpha\sim\Gamma\left(c_{0},c_{1}\right) & \quad,c_{0},c_{1}>0 & \quad,\pi\left(\alpha\right)\propto\alpha^{c_{0}-1}\exp\left(-c_{1}\alpha\right)\\
\theta_{1}\sim\Gamma\left(c_{2},c_{3}\right) & \quad,c_{2},c_{3}>0 & \quad,\pi\left(\theta_{1}\right)\propto\theta_{1}^{c_{2}-1}\exp\left(-c_{3}\theta_{1}\right)\\
\theta_{2}\sim\Gamma\left(c_{4},c_{5}\right) & \quad,c_{4},c_{5}>0 & \quad,\pi\left(\theta_{2}\right)\propto\theta_{2}^{c_{4}-1}\exp\left(-c_{5}\theta_{2}\right)\\
\theta_{3}\sim\Gamma\left(c_{6},c_{7}\right) & \quad,c_{6},c_{7}>0 & \quad,\pi\left(\theta_{3}\right)\propto\theta_{3}^{c_{6}-1}\exp\left(-c_{7}\theta_{3}\right)\\
\theta_{4}\sim\Gamma\left(c_{8},c_{9}\right) & \quad,c_{8},c_{9}>0 & \quad,\pi\left(\theta_{4}\right)\propto\theta_{4}^{c_{8}-1}\exp\left(-c_{9}\theta_{4}\right).
\end{eqnarray*}
The joint prior from (\ref{eq:GENERAL_PRIOR}) can now be written
as
\begin{equation}
\pi\left(\alpha,\theta_{1},\theta_{2},\theta_{3},\theta_{4}\right)\propto\alpha^{c_{0}-1}\theta_{1}^{c_{2}-1}\theta_{2}^{c_{4}-1}\theta_{3}^{c_{6}-1}\theta_{4}^{c_{8}-1}\exp\left(-c_{1}\alpha-c_{3}\theta_{1}-c_{5}\theta_{2}-c_{7}\theta_{3}-c_{9}\theta_{4}\right).\label{eq:GEBS_PRIOR}
\end{equation}
The resulting joint posterior, using (\ref{eq:GEBS_LIKELIHOOD}) and
(\ref{eq:GEBS_PRIOR}), is
\begin{eqnarray*}
\pi\left(\alpha,\theta_{1},\theta_{2},\theta_{3},\theta_{4}\left|\underline{x}\right.\right) & \propto & \alpha^{c_{0}-1-\sum_{i=1}^{k}r_{i}}\theta_{1}^{c_{2}-1}\theta_{2}^{c_{4}-1}\theta_{3}^{c_{6}-1}\theta_{4}^{c_{8}-1}\\
 &  & \times\exp\left(\frac{1}{\alpha^{2}}\sum_{i=1}^{k}r_{i}\right)\exp\left(-c_{1}\alpha-c_{3}\theta_{1}-c_{5}\theta_{2}-c_{7}\theta_{3}-c_{9}\theta_{4}\right)\\
 &  & \times\left[\prod_{i=1}^{k}\prod_{j=1}^{r_{i}}\frac{x_{ij}+\frac{1}{T_{i}}\exp\left(\theta_{1}+\frac{\theta_{2}}{T_{i}}+\theta_{3}V_{i}+\frac{\theta_{4}V_{i}}{T_{i}}\right)}{x_{ij}^{\frac{3}{2}}}\right]\\
 &  & \times\exp\left\{ -\frac{1}{2\alpha^{2}}\sum_{i=1}^{k}\left[T_{i}\exp\left(-\theta_{1}-\frac{\theta_{2}}{T_{i}}-\theta_{3}V_{i}-\frac{\theta_{4}V_{i}}{T_{i}}\right)\sum_{j=1}^{r_{i}}x_{ij}\right]\right.\\
 &  & \left.\quad-\frac{1}{2\alpha^{2}}\sum_{i=1}^{k}\left[\frac{1}{T_{i}}\exp\left(\theta_{1}+\frac{\theta_{2}}{T_{i}}+\theta_{3}V_{i}+\frac{\theta_{4}V_{i}}{T_{i}}\right)\sum_{j=1}^{r_{i}}\frac{1}{x_{ij}}\right]\right\} \\
 &  & \times\prod_{i=1}^{k}\left\{ \left[T_{i}\exp\left(-\theta_{1}-\frac{\theta_{2}}{T_{i}}-\theta_{3}V_{i}-\frac{\theta_{4}V_{i}}{T_{i}}\right)\right]^{\frac{r_{i}}{2}}\right.\\
 &  & \quad\times\left[1-\Phi\left(\frac{1}{\alpha}\left(\sqrt{\tau_{i}T_{i}\exp\left(-\theta_{1}-\frac{\theta_{2}}{T_{i}}-\theta_{3}V_{i}-\frac{\theta_{4}V_{i}}{T_{i}}\right)}\right.\right.\right.\\
 &  & \left.\left.\left.\left.\quad\quad-\sqrt{\frac{1}{\tau_{i}T_{i}}\exp\left(\theta_{1}+\frac{\theta_{2}}{T_{i}}+\theta_{3}V_{i}+\frac{\theta_{4}V_{i}}{T_{i}}\right)}\right)\right)\right]^{n_{i}-r_{i}}\right\} .
\end{eqnarray*}
See Smit \& Raubenheimer$^{19}$ for further details on this model.
The posterior is mathematically intractable, hence MCMC methods have
to be employed to draw posterior samples to be used for inference. 

\subsection{The GEW model}

Let $X$ be a continuous random variable that follows a Weibull distribution
with scale parameter $\alpha$ and shape parameter $\beta$ $\left(\alpha>0,\beta>0\right)$.
The PDF is then given by
\begin{eqnarray}
f\left(x\left|\alpha,\beta\right.\right) & = & \alpha\beta x^{\beta-1}\exp\left(-\alpha x^{\beta}\right),\quad x\geq0,\label{eq:PDF_WEIBULL}
\end{eqnarray}
and the reliability function by
\[
R\left(x\right)=\exp\left(-\alpha x^{\beta}\right).
\]

\noindent A common assumption in the literature is that the Weibull
scale parameter $\alpha$ is then dependent on the stress levels,
whereas the shape parameter $\beta$ is not (see, for example, Mazzuchi
\textit{et al}$^{24}$, Soyer \textit{et al}$^{8}$, Upadhyay \& Mukherjee$^{3}$).
The reparameterization of $\alpha$ given by the generalized Eyring
model, for this formulation of the Weibull model, is
\begin{eqnarray}
\alpha_{i} & = & T_{i}\exp\left(-\theta_{1}-\frac{\theta_{2}}{T_{i}}-\theta_{3}V_{i}-\frac{\theta_{4}V_{i}}{T_{i}}\right),\label{eq:GEW_ALPHA}
\end{eqnarray}
where $\theta_{1},$ $\theta_{2},$ $\theta_{3},$ and $\theta_{4}$
are unknown model parameters, and $V_{i}$ is a function of the non-thermal
stressor $S_{i}$ (Escobar \& Meeker$^{23}$). For a lifetime subjected
to the $i^{th}$ level of the stressors, it follows from (\ref{eq:PDF_WEIBULL})
and (\ref{eq:GEW_ALPHA}) that the Weibull PDF can be written as
\begin{eqnarray}
f\left(x_{i}\left|\theta_{1},\theta_{2},\theta_{3},\theta_{4},\beta\right.\right) & = & T_{i}\exp\left(-\theta_{1}-\frac{\theta_{2}}{T_{i}}-\theta_{3}V_{i}-\frac{\theta_{4}V_{i}}{T_{i}}\right)\beta x_{i}^{\beta-1}\nonumber \\
 &  & \times\exp\left[-T_{i}\exp\left(-\theta_{1}-\frac{\theta_{2}}{T_{i}}-\theta_{3}V_{i}-\frac{\theta_{4}V_{i}}{T_{i}}\right)x_{i}^{\beta}\right].\label{eq:GEW_PDF}
\end{eqnarray}
The corresponding reliability function at some time $\tau$ is given
by
\begin{equation}
R\left(\tau\right)=\exp\left[-T_{i}\exp\left(-\theta_{1}-\frac{\theta_{2}}{T_{i}}-\theta_{3}V_{i}-\frac{\theta_{4}V_{i}}{T_{i}}\right)\tau^{\beta}\right].\label{eq:GEW_RF}
\end{equation}
From (\ref{eq:GEW_PDF}) and (\ref{eq:GEW_RF}) it follows that the
likelihood function for the GEW model is given by

\begin{eqnarray}
L\left(\underline{x}\left|\theta_{1},\theta_{2},\theta_{3},\theta_{4},\beta\right.\right) & = & \beta^{\sum_{i=1}^{k}r_{i}}\exp\left(-\theta_{1}\sum_{i=1}^{k}r_{i}-\theta_{2}\sum_{i=1}^{k}\frac{r_{i}}{T_{i}}-\theta_{3}\sum_{i=1}^{k}r_{i}V_{i}-\theta_{4}\sum_{i=1}^{k}\frac{r_{i}V_{i}}{T_{i}}\right)\nonumber \\
 &  & \times\exp\left[-\sum_{i=1}^{k}\left(n_{i}-r_{i}\right)T_{i}\exp\left(-\theta_{1}-\frac{\theta_{2}}{T_{i}}-\theta_{3}V_{i}-\frac{\theta_{4}V_{i}}{T_{i}}\right)\tau_{i}^{\beta}\right]\nonumber \\
 &  & \times\exp\left[-\sum_{i=1}^{k}\sum_{j=1}^{r_{i}}T_{i}\exp\left(-\theta_{1}-\frac{\theta_{2}}{T_{i}}-\theta_{3}V_{i}-\frac{\theta_{4}V_{i}}{T_{i}}\right)x_{ij}^{\beta}\right]\left[\prod_{i=1}^{k}\prod_{j=1}^{r_{i}}T_{i}x_{ij}^{\beta-1}\right].\label{eq:GEW_LIKELIHOOD}
\end{eqnarray}
Assume that the priors on the unknown parameters $\theta_{1},$ $\theta_{2},$
$\theta_{3},$ $\theta_{4}$ and $\beta$ are independent (see, for
example, Soyer \textit{et al}$^{8}$, Upadhyay \& Mukherjee$^{3}$),
and given by
\[
\pi\left(\theta_{1},\theta_{2},\theta_{3},\theta_{4},\beta\right)=\pi\left(\theta_{1}\right)\pi\left(\theta_{2}\right)\pi\left(\theta_{3}\right)\pi\left(\theta_{4}\right)\pi\left(\beta\right).
\]
The joint posterior distribution is then given by
\begin{eqnarray*}
\pi\left(\theta_{1},\theta_{2},\theta_{3},\theta_{4},\beta\left|\underline{x}\right.\right) & \propto & L\left(\underline{x}\left|\theta_{1},\theta_{2},\theta_{3},\theta_{4},\beta\right.\right)\pi\left(\theta_{1},\theta_{2},\theta_{3},\theta_{4},\beta\right).
\end{eqnarray*}
Gamma priors are imposed on all the parameters, with
\begin{eqnarray*}
\theta_{1}\sim\Gamma(c_{10},c_{11}) & ,c_{10},c_{11}>0 & ,\pi_{2}\left(\theta_{1}\right)\propto\theta_{1}^{c_{10}-1}\exp\left(-c_{11}\theta_{1}\right)\\
\theta_{2}\sim\Gamma(c_{12},c_{13}) & ,c_{12},c_{13}>0 & ,\pi_{2}\left(\theta_{2}\right)\propto\theta_{2}^{c_{12}-1}\exp\left(-c_{13}\theta_{2}\right)\\
\theta_{3}\sim\Gamma(c_{14},c_{15}) & ,c_{14},c_{15}>0 & ,\pi_{2}\left(\theta_{3}\right)\propto\theta_{3}^{c_{14}-1}\exp\left(-c_{15}\theta_{3}\right)\\
\theta_{4}\sim\Gamma(c_{16},c_{17}) & ,c_{16},c_{17}>0 & ,\pi_{2}\left(\theta_{4}\right)\propto\theta_{4}^{c_{16}-1}\exp\left(-c_{17}\theta_{4}\right)\\
\beta\sim\Gamma(c_{18},c_{19}) & ,c_{18},c_{19}>0 & ,\pi_{2}\left(\beta\right)\propto\beta^{c_{18}-1}\exp\left(-c_{19}\beta\right).
\end{eqnarray*}
The joint prior for the GEW model is then given by
\begin{equation}
\pi_{2}\left(\theta_{1},\theta_{2},\theta_{3},\theta_{4},\beta\right)\propto\theta_{1}^{c_{10}-1}\theta_{2}^{c_{12}-1}\theta_{3}^{c_{14}-1}\theta_{4}^{c_{16}-1}\beta^{c_{18}-1}\exp\left(-c_{11}\theta_{1}-c_{13}\theta_{2}-c_{15}\theta_{3}-c_{17}\theta_{4}-c_{19}\beta\right),\label{eq:GEW2_PRIORS}
\end{equation}
and, using (\ref{eq:GEW_LIKELIHOOD}) and (\ref{eq:GEW2_PRIORS}),
the joint posterior is given by
\begin{eqnarray*}
\pi_{2}\left(\theta_{1},\theta_{2},\theta_{3},\theta_{4},\beta\left|\underline{x}\right.\right) & \propto & \theta_{1}^{c_{10}-1}\theta_{2}^{c_{12}-1}\theta_{3}^{c_{14}-1}\theta_{4}^{c_{16}-1}\beta^{c_{18}-1}\exp\left(-c_{11}\theta_{1}-c_{13}\theta_{2}-c_{15}\theta_{3}-c_{17}\theta_{4}-c_{19}\beta\right)\\
 &  & \times\beta^{\sum_{i=1}^{k}r_{i}}\exp\left(-\theta_{1}\sum_{i=1}^{k}r_{i}-\theta_{2}\sum_{i=1}^{k}\frac{r_{i}}{T_{i}}-\theta_{3}\sum_{i=1}^{k}r_{i}V_{i}-\theta_{4}\sum_{i=1}^{k}\frac{r_{i}V_{i}}{T_{i}}\right)\\
 &  & \times\exp\left[-\sum_{i=1}^{k}\left(n_{i}-r_{i}\right)T_{i}\exp\left(-\theta_{1}-\frac{\theta_{2}}{T_{i}}-\theta_{3}V_{i}-\frac{\theta_{4}V_{i}}{T_{i}}\right)\tau_{i}^{\beta}\right]\\
 &  & \times\exp\left[-\sum_{i=1}^{k}\sum_{j=1}^{r_{i}}T_{i}\exp\left(-\theta_{1}-\frac{\theta_{2}}{T_{i}}-\theta_{3}V_{i}-\frac{\theta_{4}V_{i}}{T_{i}}\right)x_{ij}^{\beta}\right]\left[\prod_{i=1}^{k}\prod_{j=1}^{r_{i}}T_{i}x_{ij}^{\beta-1}\right].
\end{eqnarray*}
See Smit$^{18}$ for further details on this model. Due to the complexity
of the posterior, MCMC methods are used to draw posterior samples
to base inferences on. 

\section{Estimating the Marginal Likelihood}

There are various methods that can be used to estimate the marginal
likelihood. In this section, we focus on methods that can easily estimate
Bayes factors from the output of an MCMC algorithm. This includes
the simple Monte Carlo estimator, the Laplace-Metropolis estimator,
the harmonic mean estimator, and using the predictive posterior density
to estimate the posterior Bayes factors.

\subsection{Simple Monte Carlo Estimator}

Since the marginal likelihood for a model $m_{i}$ is given by
\[
f\left(\underline{x}\left|m_{i}\right.\right)=\int f\left(\underline{x}\left|\bm{\theta_{m_{i}}},m_{i}\right.\right)f\left(\bm{\theta_{m_{i}}}\left|m_{i}\right.\right)d\bm{\theta_{m_{i}}},
\]
a straightforward and simple estimate is provided by the Monte Carlo
integration estimate
\[
\hat{f}_{MC}\left(\underline{x}\left|m_{i}\right.\right)=\frac{1}{N}\sum_{t=1}^{N}f\left(\underline{x}\left|\bm{\theta_{m_{i}}^{*(t)}},m_{i}\right.\right),
\]
where $\bm{\theta_{m_{i}}^{*(1)}},\bm{\theta_{m_{i}}^{*(2)}},\dots,\bm{\theta_{m_{i}}^{*(N)}}$
are samples from the prior distribution of model $m_{i}$. According
to Kass \& Raftery$^{10}$ this estimator can be very inefficient,
particularly when the prior distribution and posterior distribution
significantly differ. An example of where a considerable difference
in the prior and posterior distributions can be expected, is when
a flat prior is used (Ntzoufras$^{11}$). In such a case, very small
likelihood values will be produced for most of the samples $\bm{\theta_{m_{i}}^{*(t)}}$
and the estimate will be dominated by only a few large likelihood
values.

\subsection{Laplace Approximation}

A widely used approximation for the marginal likelihood is the Laplace
approximation. The approximation is given by
\[
f\left(\underline{x}\left|m_{i}\right.\right)\approx\left(2\pi\right)^{\frac{d_{m_{i}}}{2}}\left|\bm{\tilde{\varSigma}_{m_{i}}}\right|^{\frac{1}{2}}f\left(\underline{x}\left|\bm{\tilde{\theta}_{m_{i}}},m_{i}\right.\right)f\left(\bm{\tilde{\theta}_{m_{i}}}\left|m_{i}\right.\right),
\]
where $d_{m_{i}}$ is the number of parameters for model $m_{i}$,
$\bm{\tilde{\theta}_{m_{i}}}$ is the posterior mode of the parameters
for model $m_{i}$, and $\bm{\tilde{\varSigma}_{m_{i}}}=\left[-\bm{H_{m}}\left(\bm{\tilde{\theta}_{m_{i}}}\right)\right]^{-1}$.
$\bm{H_{m}}\left(\bm{\tilde{\theta}_{m_{i}}}\right)$ is the Hessian
matrix of second derivatives for the log of the posterior density,
$\ln\left[f\left(\bm{\theta_{m_{i}}}\left|\underline{x},m_{i}\right.\right)\right]$,
evaluated at the posterior mode $\bm{\tilde{\theta}_{m_{i}}}$. Kass
\& Raftery$^{10}$ state that the Laplace approximation works well
for symmetric likelihood functions and for a parameter vector of moderate
dimensionality.

Raftery$^{25}$ and Lewis \& Raftery$^{26}$ propose an extension
of the Laplace approximation, called the Laplace-Metropolis estimator,
which avoids the analytical calculation of $\bm{\tilde{\varSigma}_{m_{i}}}$
and $\bm{\tilde{\theta}_{m_{i}}}$. The posterior mean and variance-covariance
matrix of the posterior sample $\bm{\theta_{m_{i}}^{(1)}},\bm{\theta_{m_{i}}^{(2)}},\dots,\bm{\theta_{m_{i}}^{(N)}}$,
generated from an MCMC algorithm, are used to estimate $\bm{\tilde{\theta}_{m_{i}}}$
and $\bm{\tilde{\varSigma}_{m_{i}}}$, respectively. The Laplace-Metropolis
estimator is given by
\[
\hat{f}_{LM}\left(\underline{x}\left|m_{i}\right.\right)=\left(2\pi\right)^{\frac{d_{m_{i}}}{2}}\left|\bm{S_{m_{i}}}\right|^{\frac{1}{2}}f\left(\underline{x}\left|\bm{\overline{\theta}_{m_{i}}},m_{i}\right.\right)f\left(\bm{\overline{\theta}_{m_{i}}}\left|m_{i}\right.\right),
\]
where $\bm{\overline{\theta}_{m_{i}}}=\frac{1}{N}\sum_{t=1}^{N}\bm{\theta_{m_{i}}^{(t)}}$
and $\bm{S_{m_{i}}}$ is a weighted variance matrix estimate (see,
Lewis \& Raftery$^{26}$, for further details).

Ntzoufras$^{11}$ explains how the Laplace-Metropolis estimator can
be calculated from an MCMC output in OpenBUGS as follows:
\begin{enumerate}
\item Implement the model $m_{i}$ in OpenBUGS and produce posterior samples
for the parameters of interest via an MCMC algorithm.
\item From the MCMC samples calculate the following estimates:
\begin{itemize}
\item $\bm{\overline{\theta}_{m_{i}}}=\left(\overline{\theta}_{1},\overline{\theta}_{2},\dots,\overline{\theta}_{d_{m_{i}}}\right)$,
which are the posterior means of the parameters of interest.
\item $\bm{\bm{s}_{\theta_{m_{i}}}}=\left(s_{\theta_{1}},s_{\theta_{2}},\dots,s_{\theta_{d_{m_{i}}}}\right)$,
which are the posterior standard deviations of the parameters of interest.
\item $\bm{R}_{\bm{\theta_{m_{i}}}}$, which is the posterior correlation
matrix between the parameters of interest.
\end{itemize}
\item Calculate the expression
\[
\ln\hat{f}_{LM}\left(\underline{x}\left|m_{i}\right.\right)=\frac{1}{2}d_{m_{i}}\ln\left(2\pi\right)+\frac{1}{2}\ln\left|\bm{R}_{\bm{\theta_{m_{i}}}}\right|+\sum_{l=1}^{d_{m_{i}}}\ln s_{\theta_{l}}+\sum_{k=1}^{n}\ln f\left(x_{k}\left|\bm{\overline{\theta}_{m_{i}}},m_{i}\right.\right)+\ln f\left(\bm{\overline{\theta}_{m_{i}}}\left|m_{i}\right.\right),
\]
and simply take $e^{\ln\hat{f}_{LM}\left(\underline{x}\left|m_{i}\right.\right)}$
to get the Laplace-Metropolis estimate for the marginal likelihood.
\end{enumerate}

\subsection{Harmonic Mean Estimator}

The harmonic mean estimator for the marginal likelihood, introduced
in Newton \& Raftery$^{27}$, is given by 
\[
\hat{f}_{HM}\left(\underline{x}\left|m_{i}\right.\right)=\left[\frac{1}{N}\sum_{t=1}^{N}\left(f\left(\underline{x}\left|\bm{\theta_{m_{i}}^{(t)}},m_{i}\right.\right)\right)^{-1}\right]^{-1},
\]
where $\bm{\theta_{m_{i}}^{(1)}},\bm{\theta_{m_{i}}^{(2)}},\dots,\bm{\theta_{m_{i}}^{(N)}}$
are posterior samples generated by an MCMC method or other sampling
technique. According to Newton \& Raftery$^{27}$, $\hat{f}_{HM}\left(\underline{x}\left|m_{i}\right.\right)$
converges almost surely to $f\left(\underline{x}\left|m_{i}\right.\right)$,
but the harmonic mean estimator does not, in general, satisfy a Gaussian
central limit theorem. This estimator is shown to be unstable and
sensitive to small likelihood values, in Raftery$^{25}$ and Raftery
\textit{et al}.$^{28}$, but it is simple to calculate. The harmonic
mean estimator is simulation-consistent and unbiased, but can have
infinite variance resulting in unstable behaviour.

A generalized harmonic mean estimator, which is an unbiased, consistent
and more stable estimator for the marginal likelihood, is presented
in Gelfand \& Dey$^{29}$. This estimator, however, requires the specification
of an importance density, which must be carefully chosen and be relatively
close to the posterior. Raftery \textit{et al}.$^{28}$ present two
more methods for stabilising the harmonic mean estimator, and improvements
on the harmonic mean estimator via Lebesgue integration is presented
in Weinberg$^{30}$.

\subsection{Posterior Predictive Density Estimate}

A variation on the traditional Bayes factor is the posterior Bayes
factor, introduced in Aitkin$^{15}$. The posterior Bayes factor of
model $m_{i}$ versus model $m_{j}$ is based on the ratio of the
PPDs of these models, for the observed data, and is given by
\[
B_{ij}=\frac{f\left(\underline{x}\left|\underline{x},m_{i}\right.\right)}{f\left(\underline{x}\left|\underline{x},m_{j}\right.\right)},
\]
where $f\left(\underline{x}\left|\underline{x},m_{g}\right.\right)$,
$g=i,j$, is the PPD of model $m_{g}$, evaluated at the observed
data. The PPD can be easily estimated by the posterior mean of the
likelihood for the posterior samples obtained from an MCMC algorithm
(Ntzoufras$^{11}$). This estimate is given by
\[
\hat{f}_{PPD}\left(\underline{x}\left|\underline{x},m_{i}\right.\right)=\frac{1}{N}\sum_{t=1}^{N}f\left(\underline{x}\left|\bm{\theta_{m_{i}}^{(t)}},m_{i}\right.\right),
\]
where $\bm{\theta_{m_{i}}^{(1)}},\bm{\theta_{m_{i}}^{(2)}},\dots,\bm{\theta_{m_{i}}^{(N)}}$
are posterior samples for the model parameters.

The use of posterior Bayes factors has been criticized due the double
use of the data, which is also the case in the construction of the
effective number of parameters $p_{D}$, used as a complexity measure
in the DIC. Posterior Bayes factors can, however, support more complex
models than traditional Bayes factors.

\section{Application\label{sec:BF_APPLICATION}}

An ALT data set from ReliaSoft$^{31}$ is used in this application.
The data represent failure times for a certain device under accelerated
temperature and relative humidity stressors. The normal use conditions
are a temperature of $T_{u}=313K$ and a relative humidity of $S_{u}=0.5$.
The data set, displayed in Table \ref{Flo:BF_DATA}, contains failure
data on 21 devices that were tested, where testing continued until
all the devices had failed. The combined effect of these stressors
must be investigated by a dual-stress acceleration model, such as
the generalized Eyring model.

The ALT models utilized in this applications are the GEW model and
the $\text{GEBS}$ model. The prior specifications for these models
are given in Table \ref{Flo:BF_PARMS}. Three choices of hyperparameters
are used for each model, in order to investigate model selection.
The models in this application are denoted by $\text{GE}\text{W}_{BF1}$,
$\text{GE}\text{W}_{BF2}$, $\text{GE}\text{W}_{BF3}$, $\text{GEB}\text{S}_{BF1}$,
$\text{GEB}\text{S}_{BF2}$, and $\text{GEB}\text{S}_{BF3}$. Flat
gamma priors are imposed on the $\text{GE}\text{W}_{BF1}$ and $\text{GEB}\text{S}_{BF1}$
models. Subjective gamma priors with mean $5$ and different variances
are used for the $\text{GE}\text{W}_{BF2}$ and $\text{GE}\text{W}_{BF3}$,
as well as for the $\text{GEB}\text{S}_{BF2}$ and $\text{GEB}\text{S}_{BF3}$
models. These subjective prior are chosen such that the DIC values
for the models differ enough to illustrate more meaningful model selection
conclusions.

Since no prior information regarding the models are available, the
prior model probabilities are set equal. This allows us to perform
model selection by using only the Bayes factors. The models are implemented
in OpenBUGS to generate posterior samples to base inference on. A
single Markov chain is initiated for each model, with a burn-in of
50000 iterations, after which 200000 samples are obtained. Trace plots
and the modified Gelman-Rubin statistic, proposed by Brooks \& Gelman$^{32}$,
are used to verify that the Markov chains have converged before the
burn-in iterations end. The Monte Carlo error is less than 5\% of
the sample standard deviation for all parameters in these models,
indicating that enough samples have been generated.

\begin{sidewaysfigure}
\begin{centering}
\includegraphics[scale=0.28]{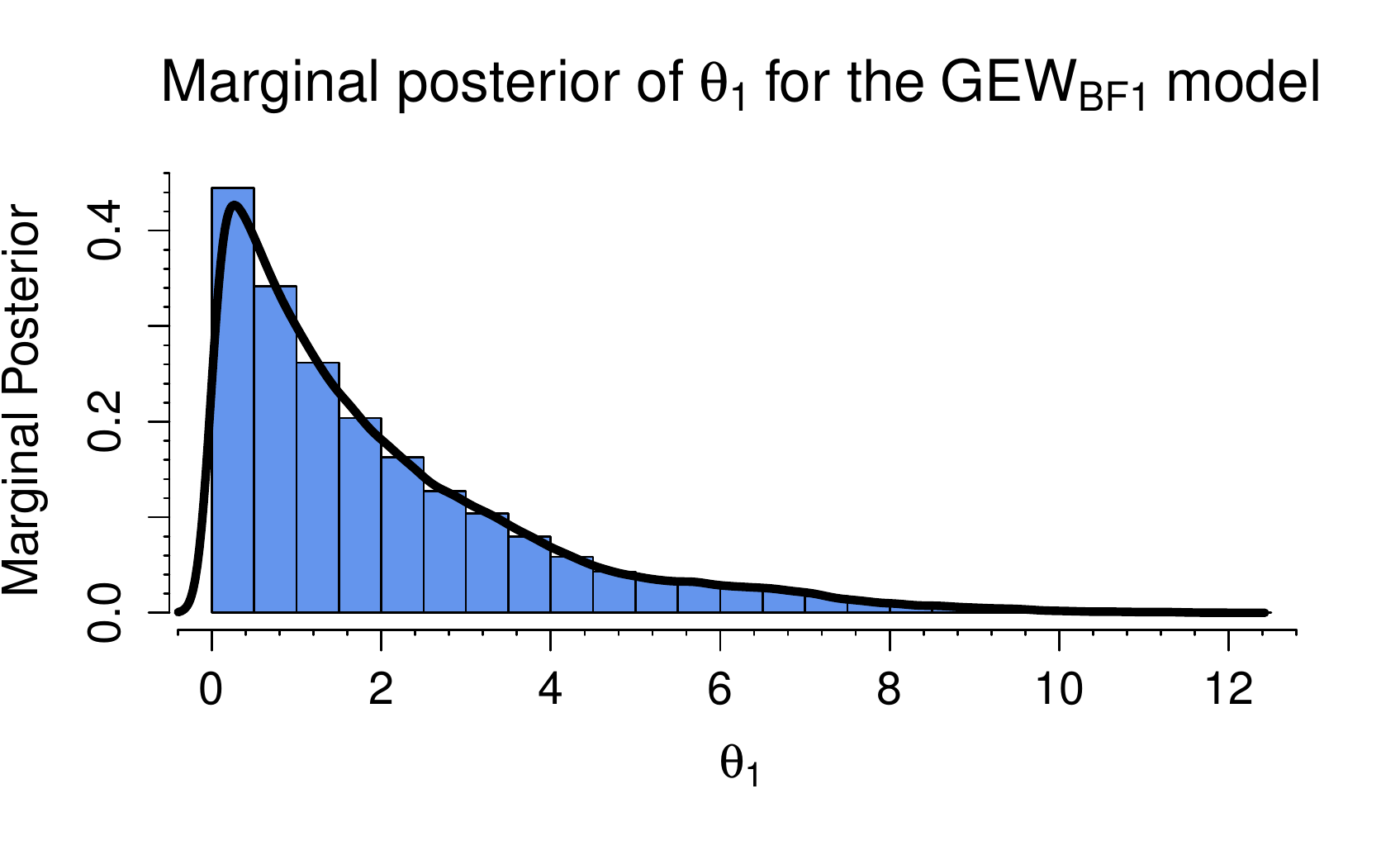}\includegraphics[scale=0.28]{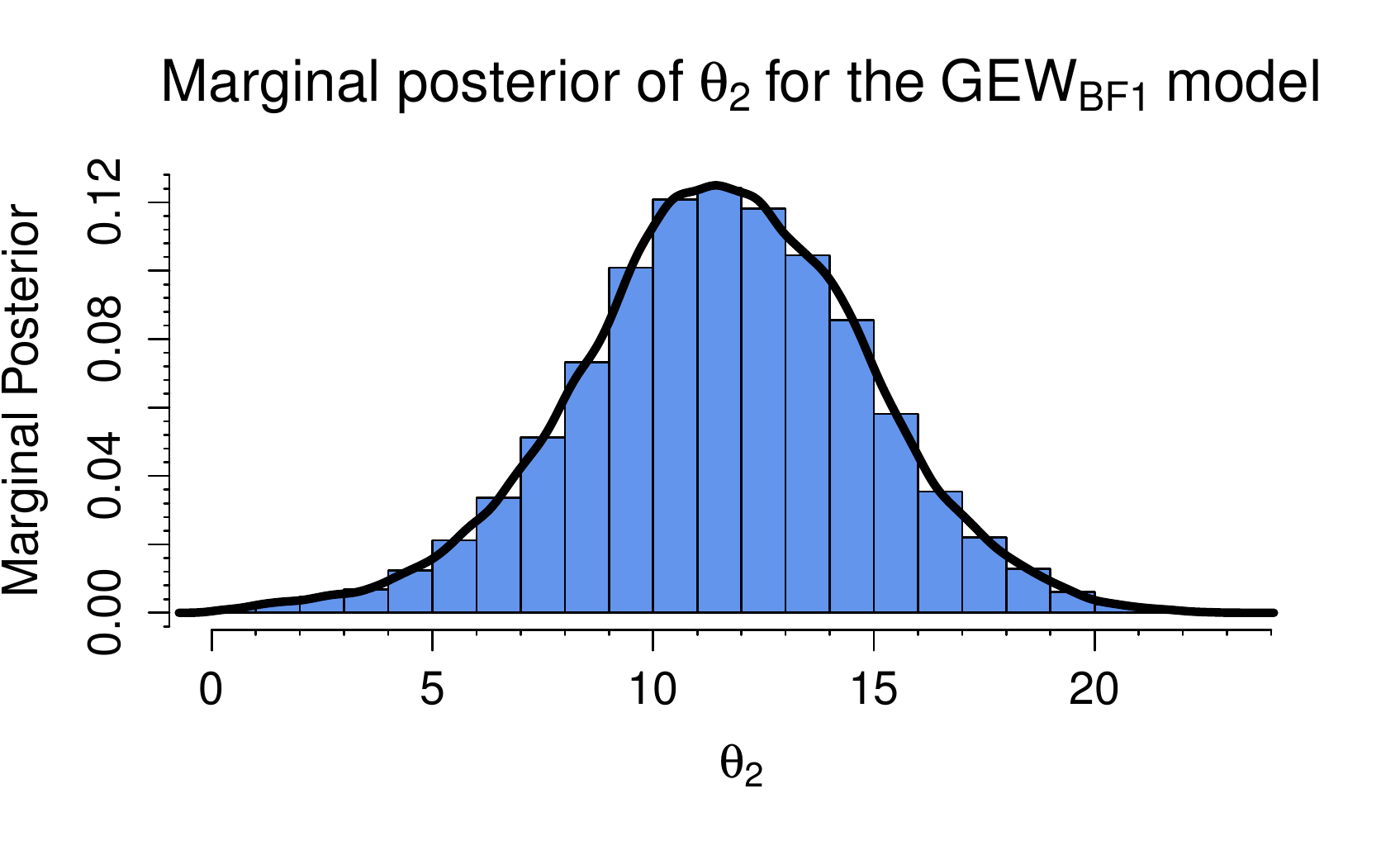}\includegraphics[scale=0.28]{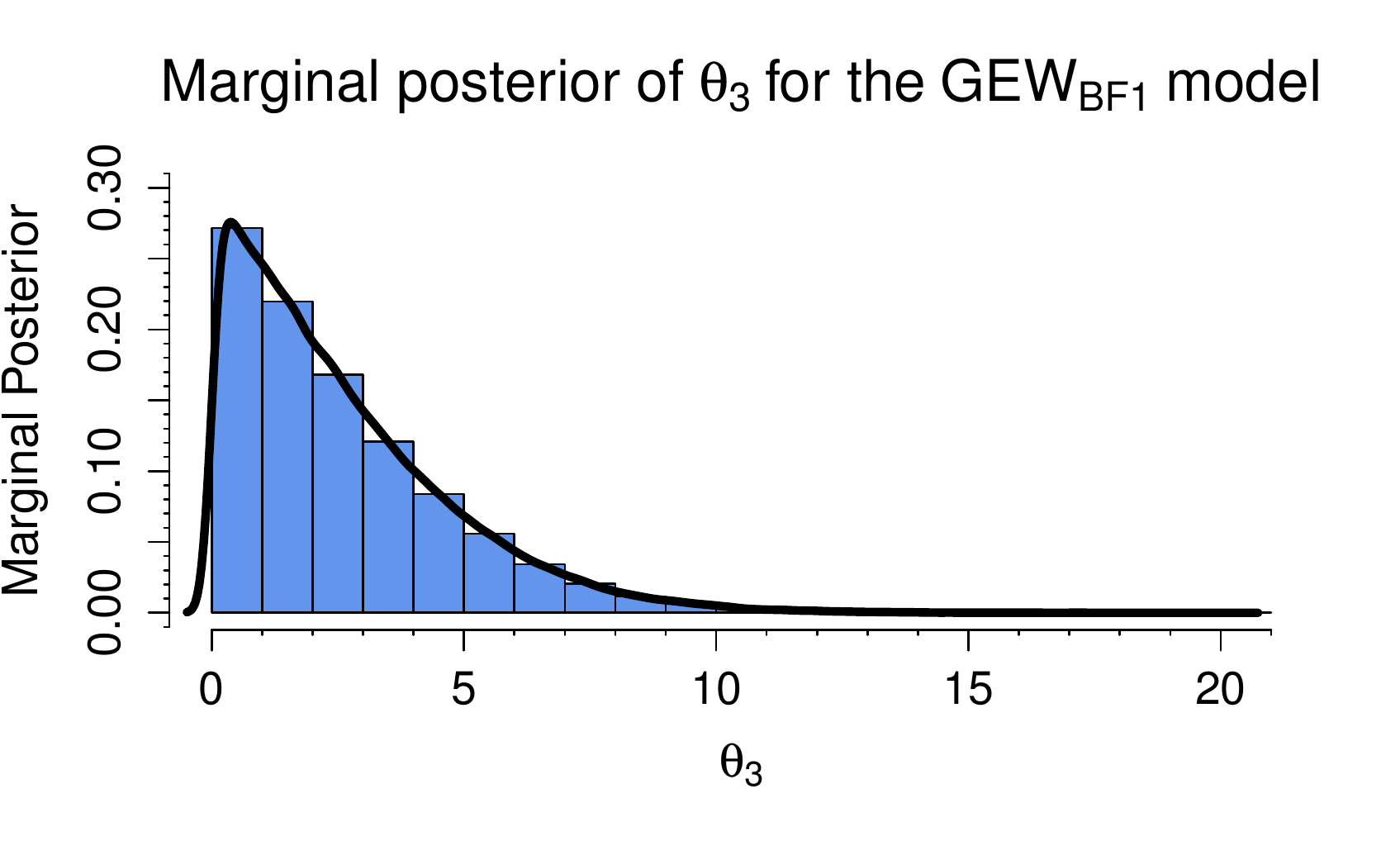}\includegraphics[scale=0.28]{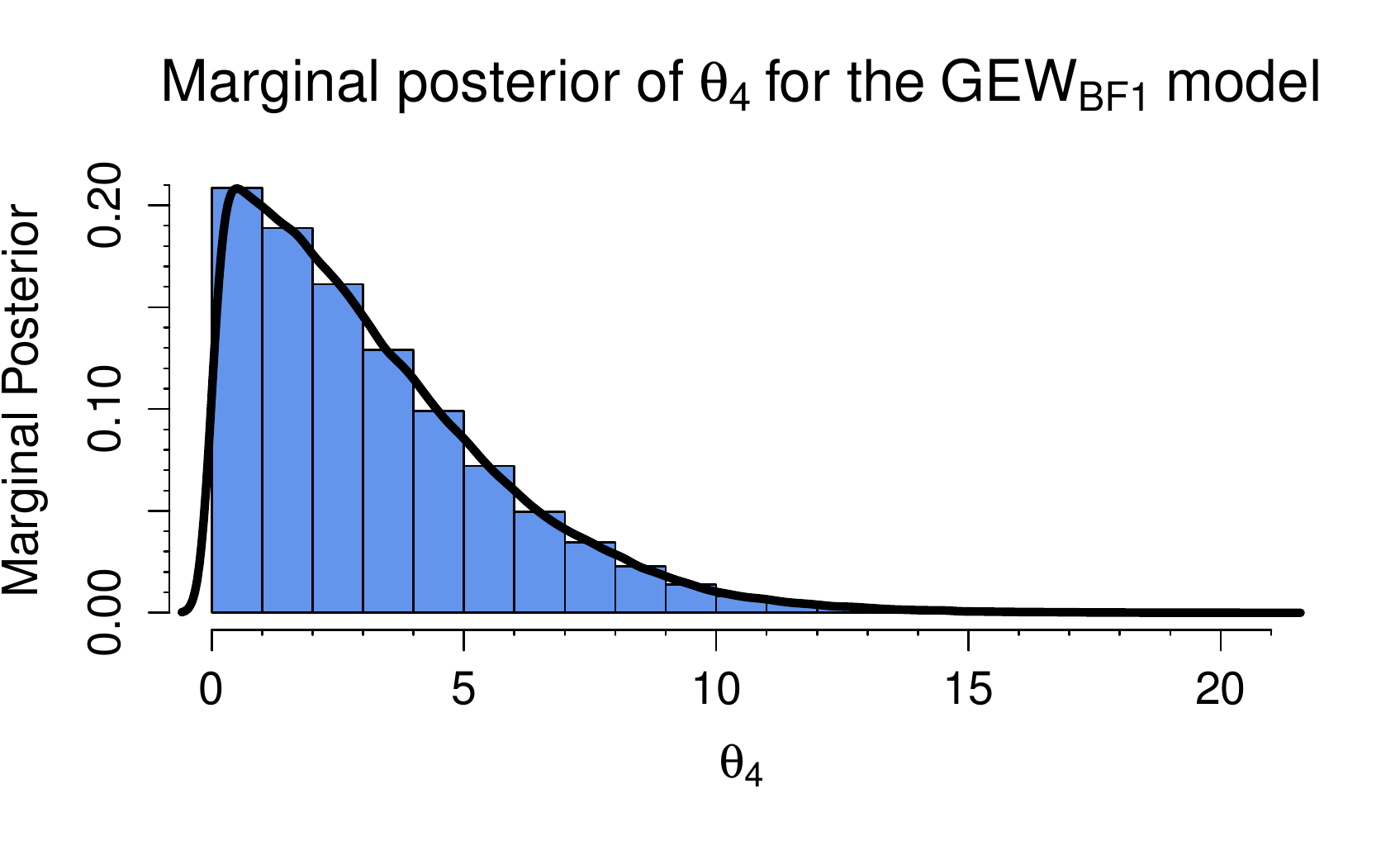}\includegraphics[scale=0.28]{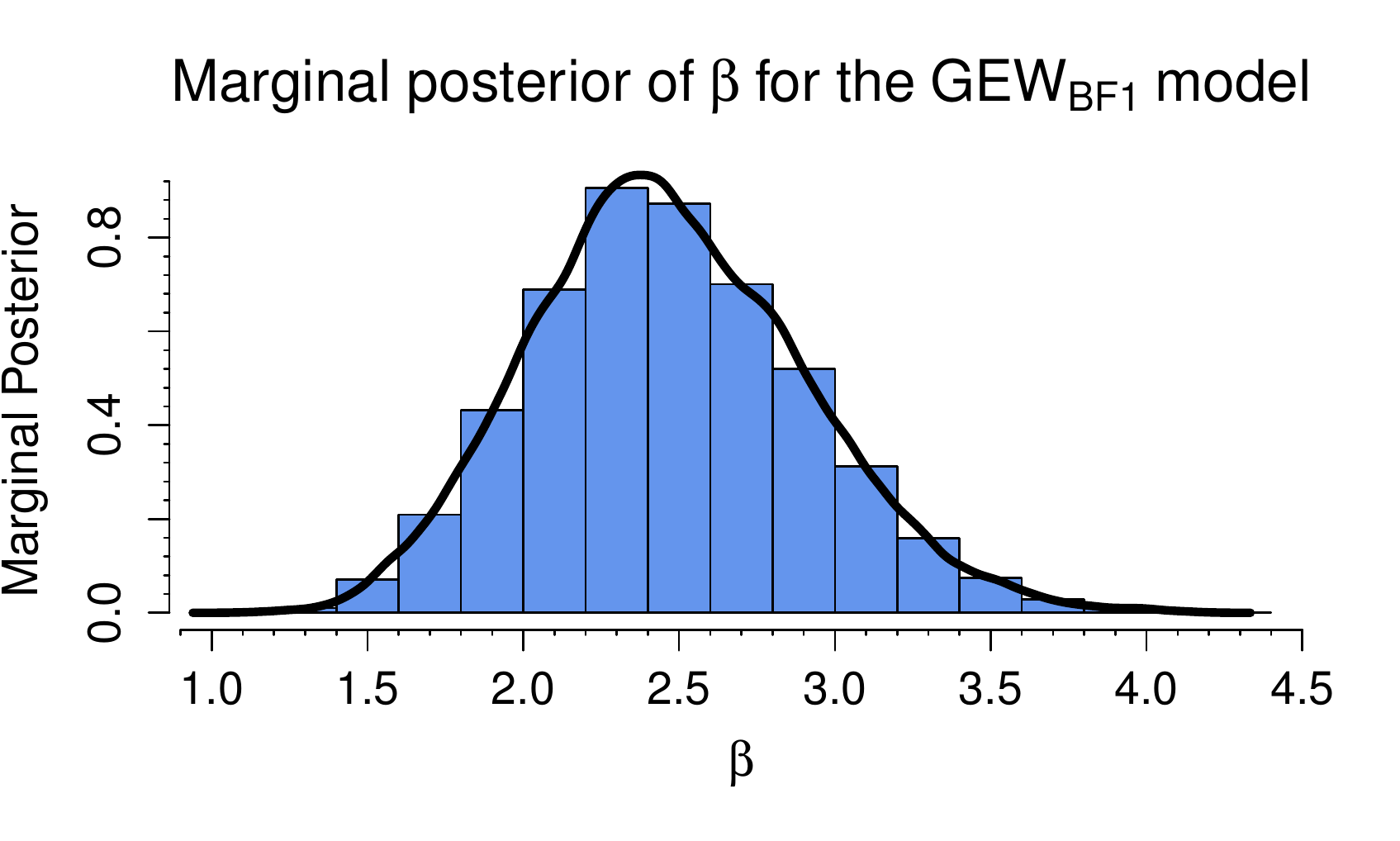}
\par\end{centering}
\begin{centering}
\includegraphics[scale=0.28]{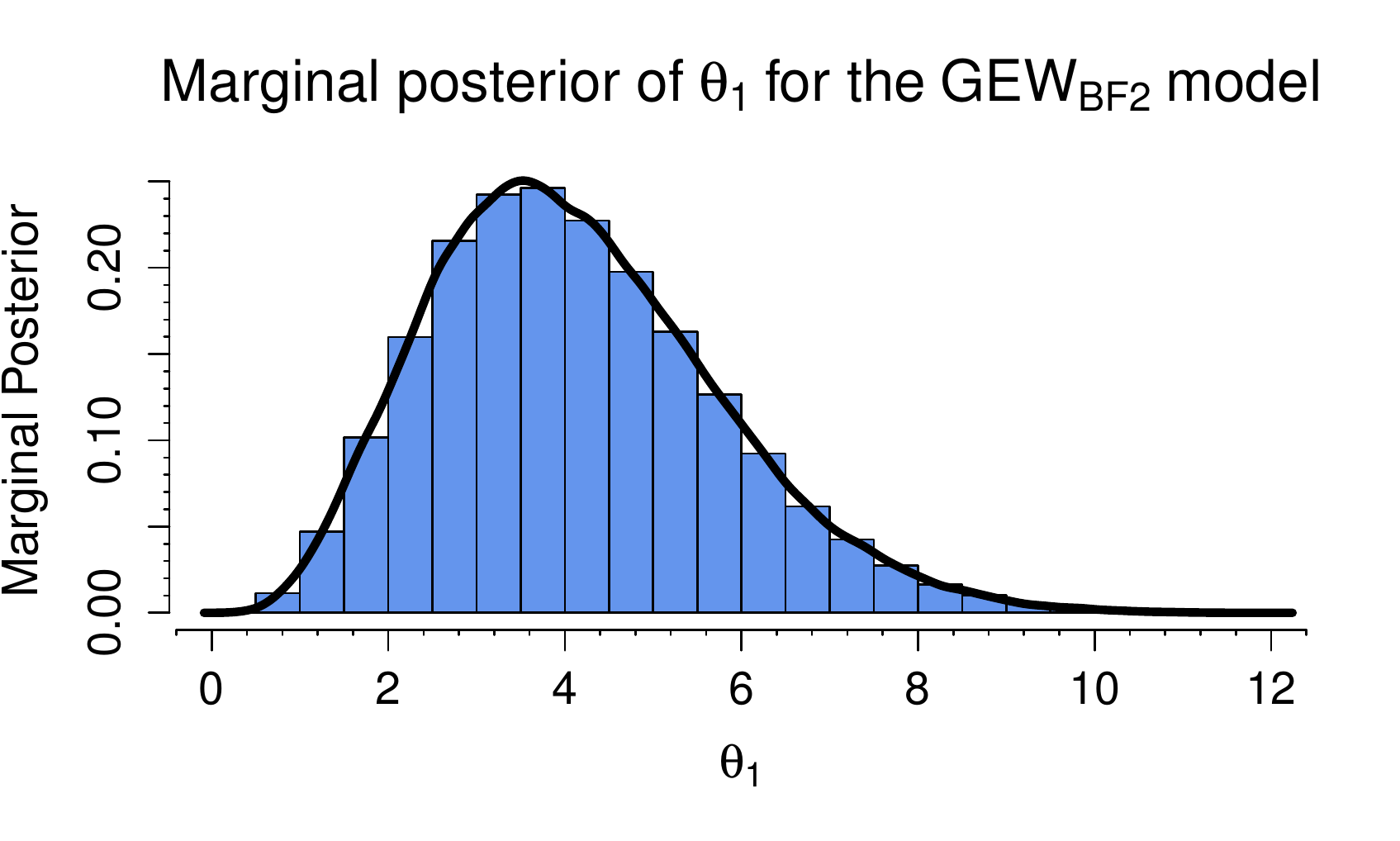}\includegraphics[scale=0.28]{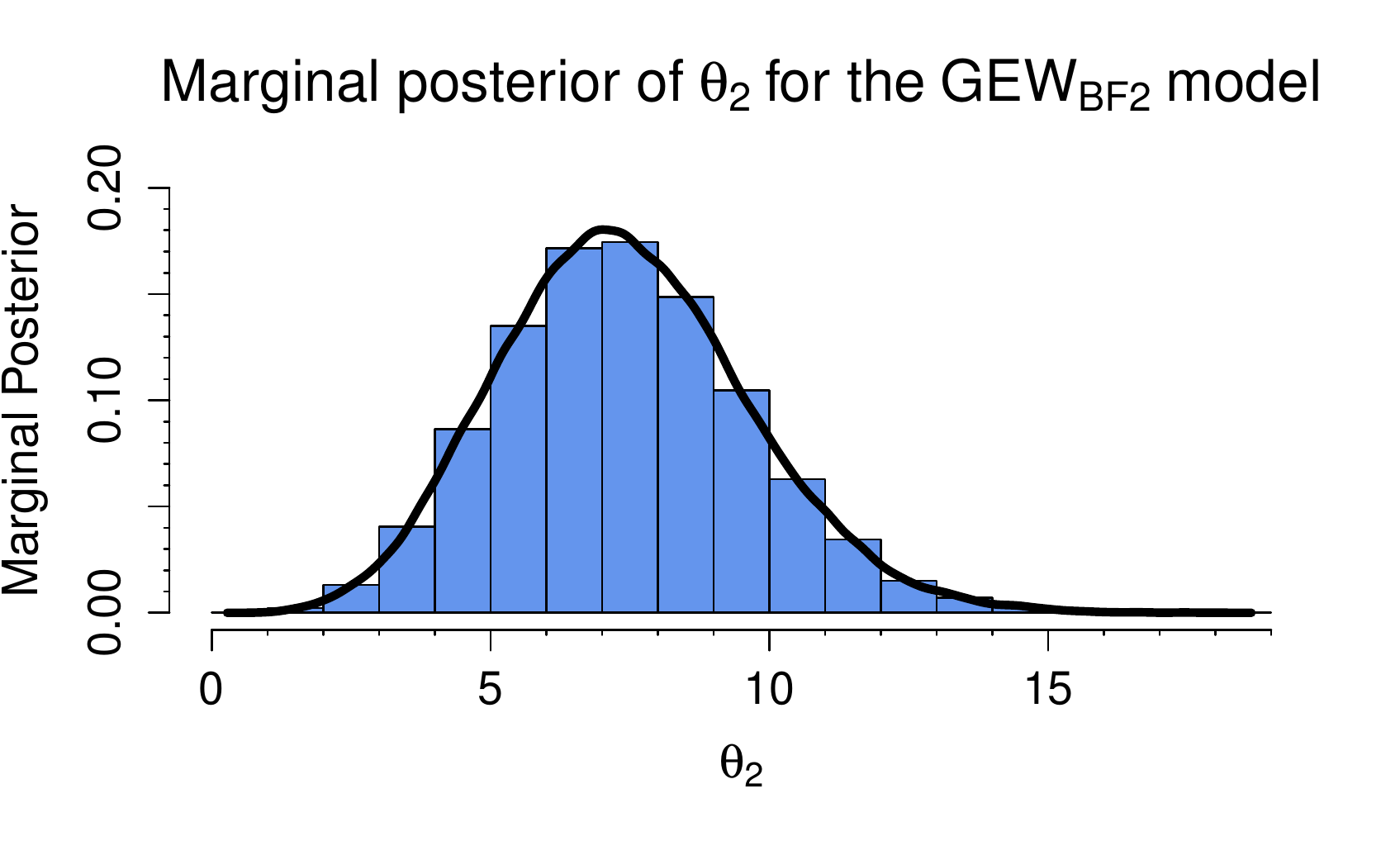}\includegraphics[scale=0.28]{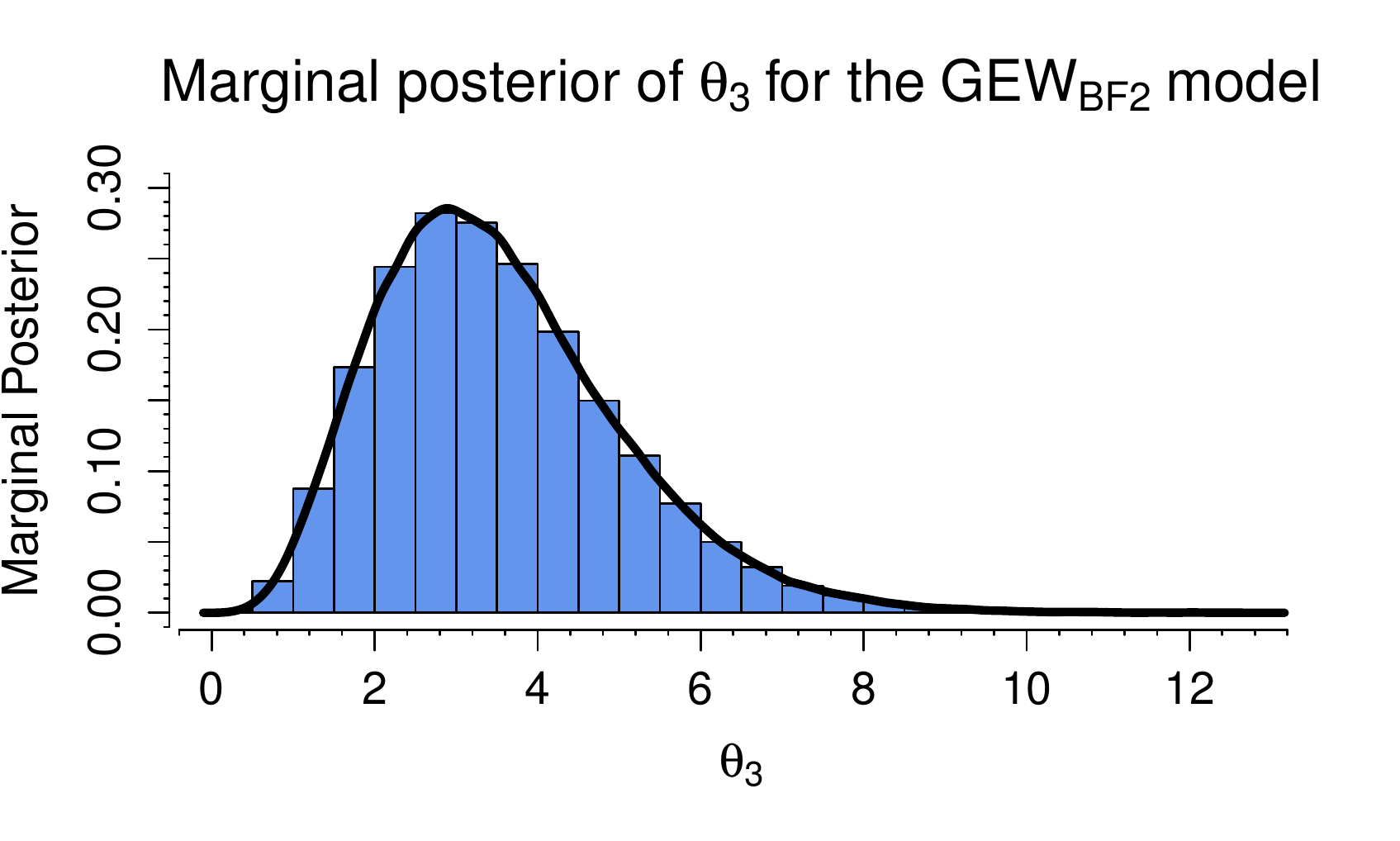}\includegraphics[scale=0.28]{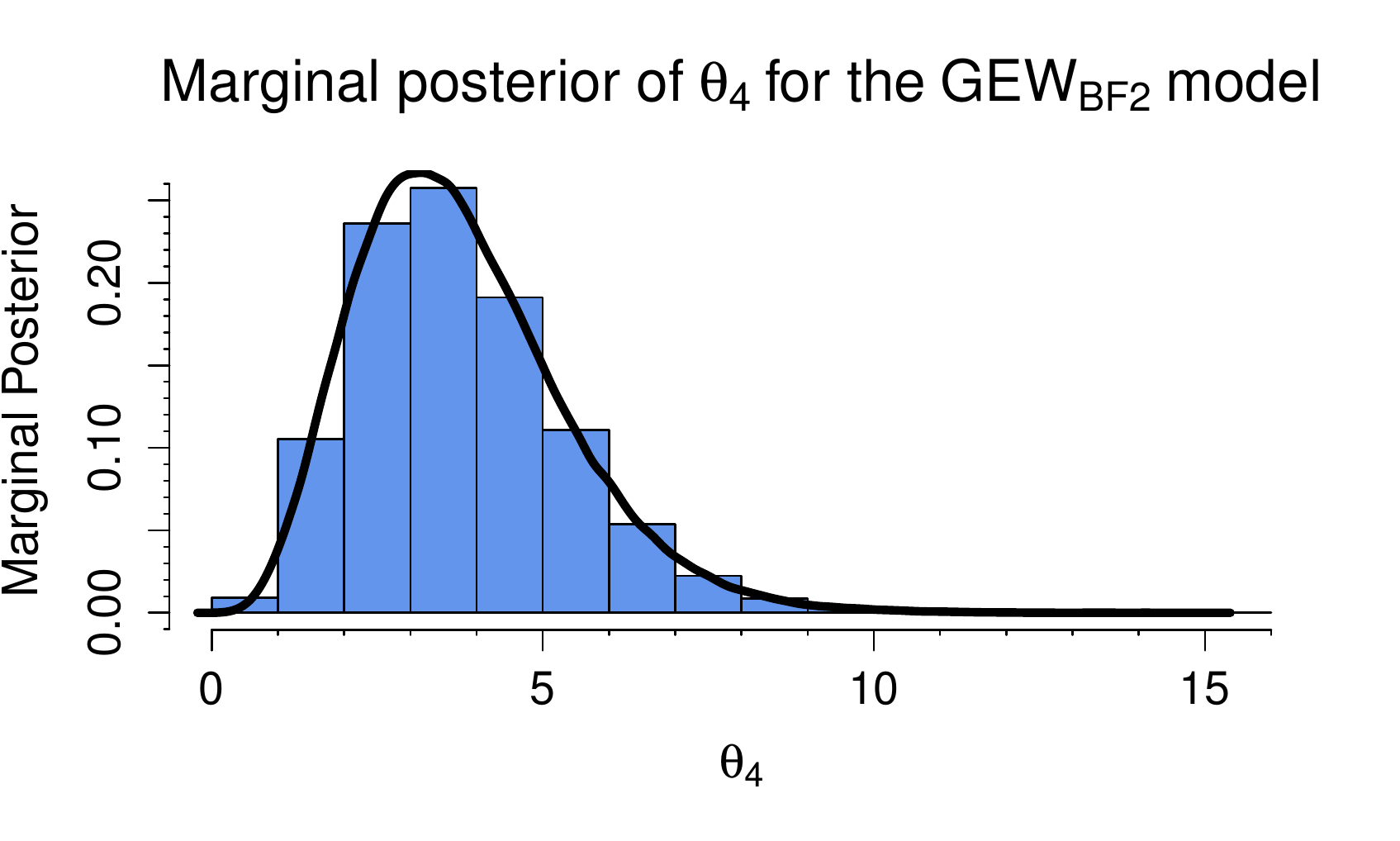}\includegraphics[scale=0.28]{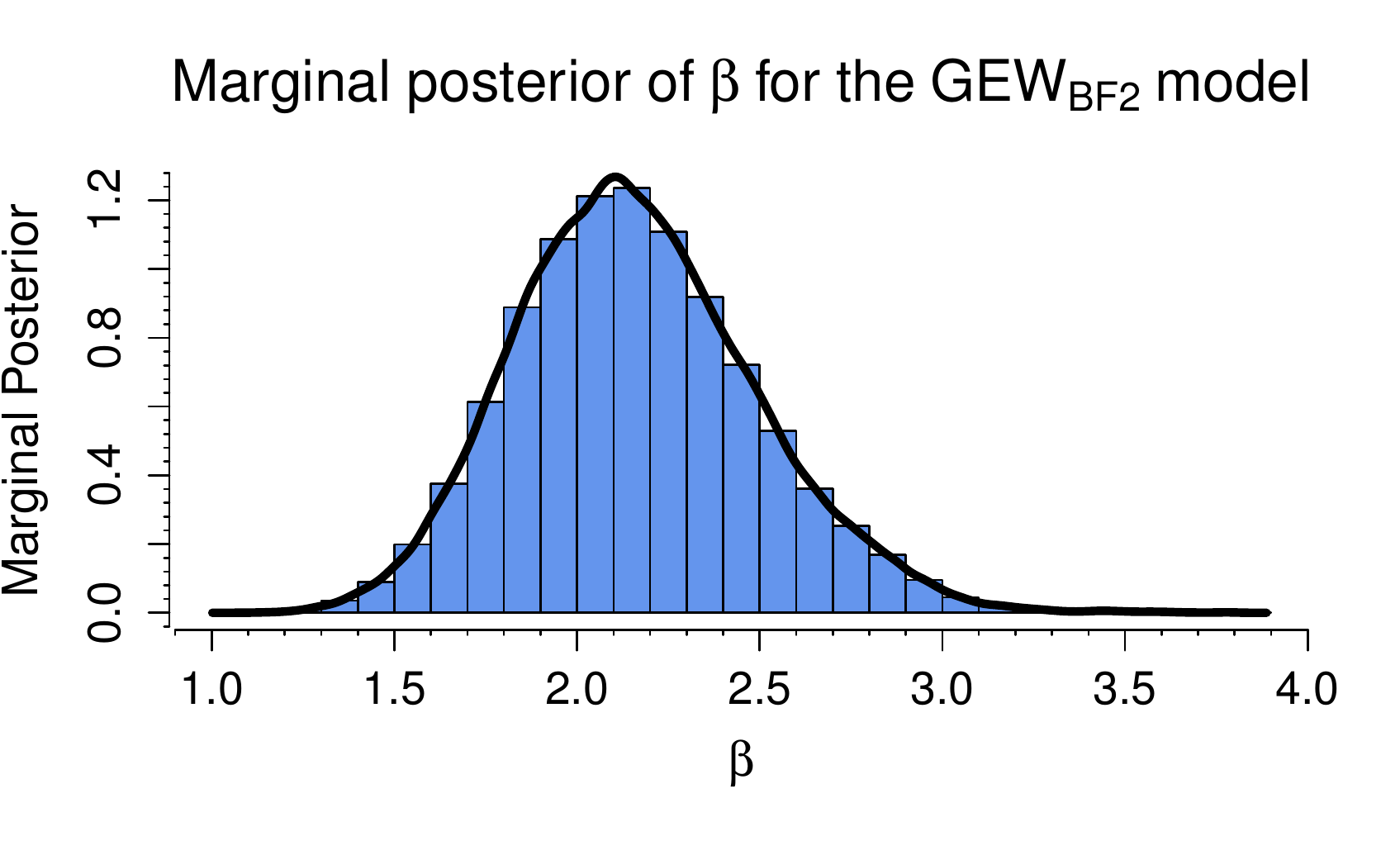}
\par\end{centering}
\begin{centering}
\includegraphics[scale=0.28]{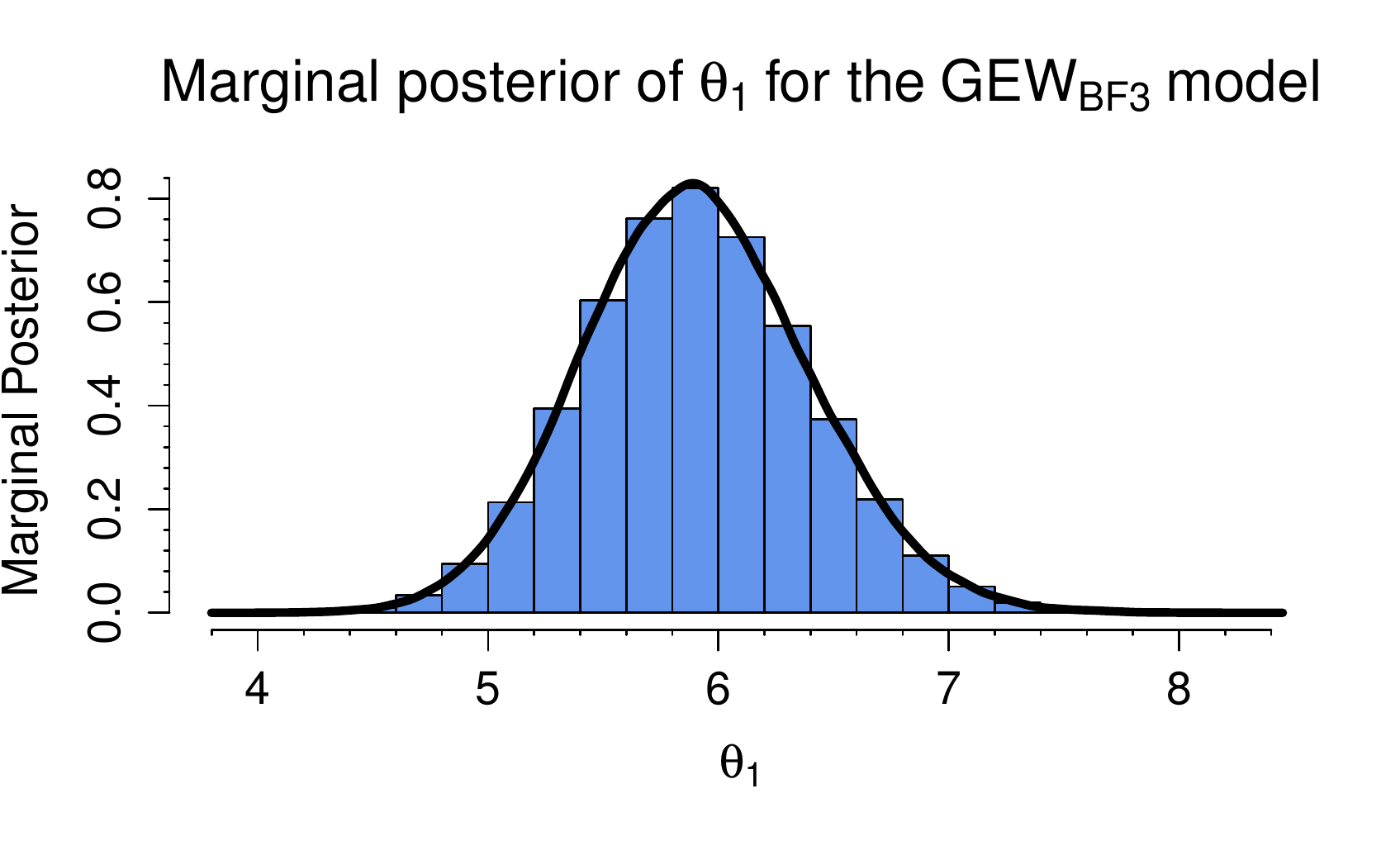}\includegraphics[scale=0.28]{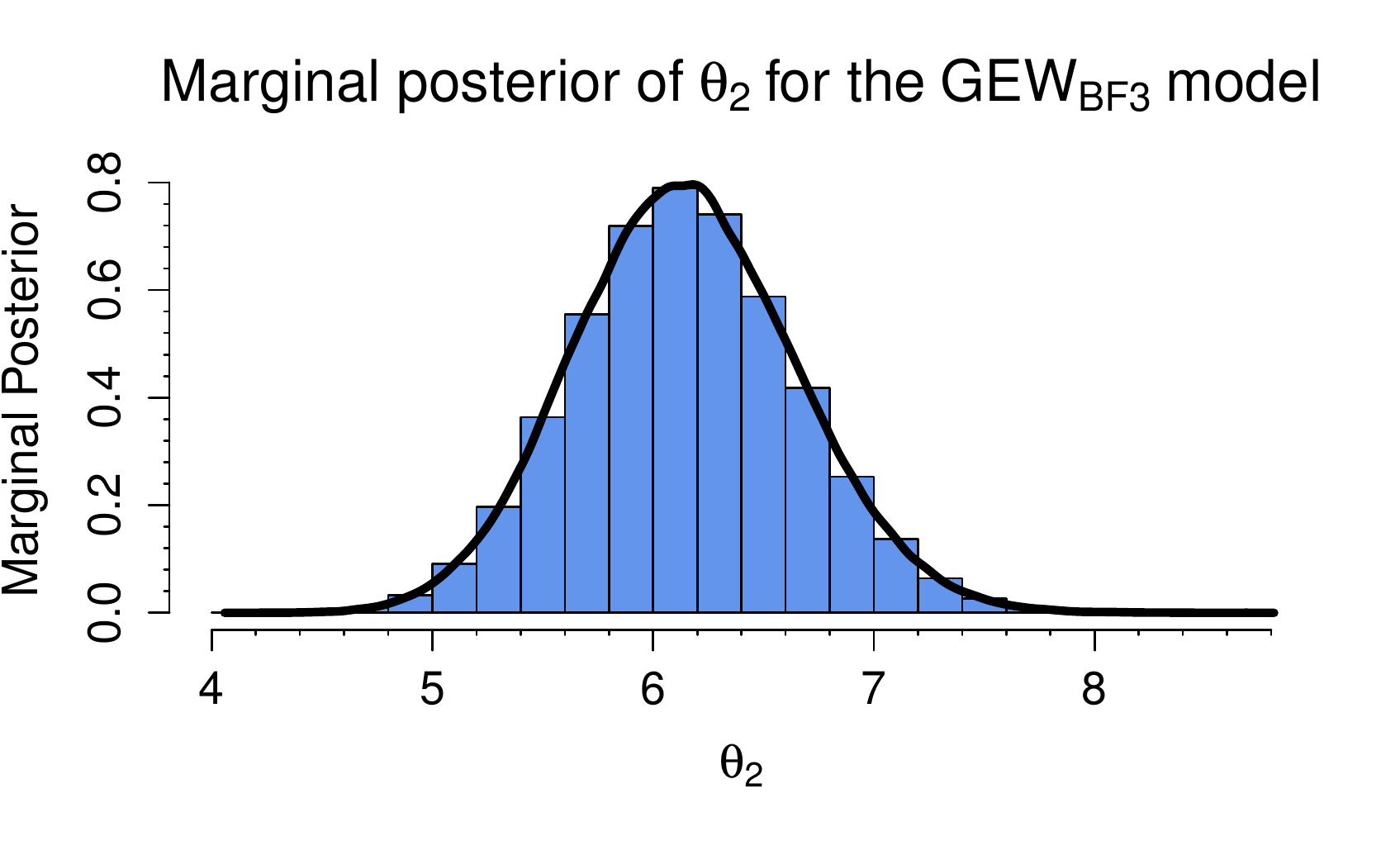}\includegraphics[scale=0.28]{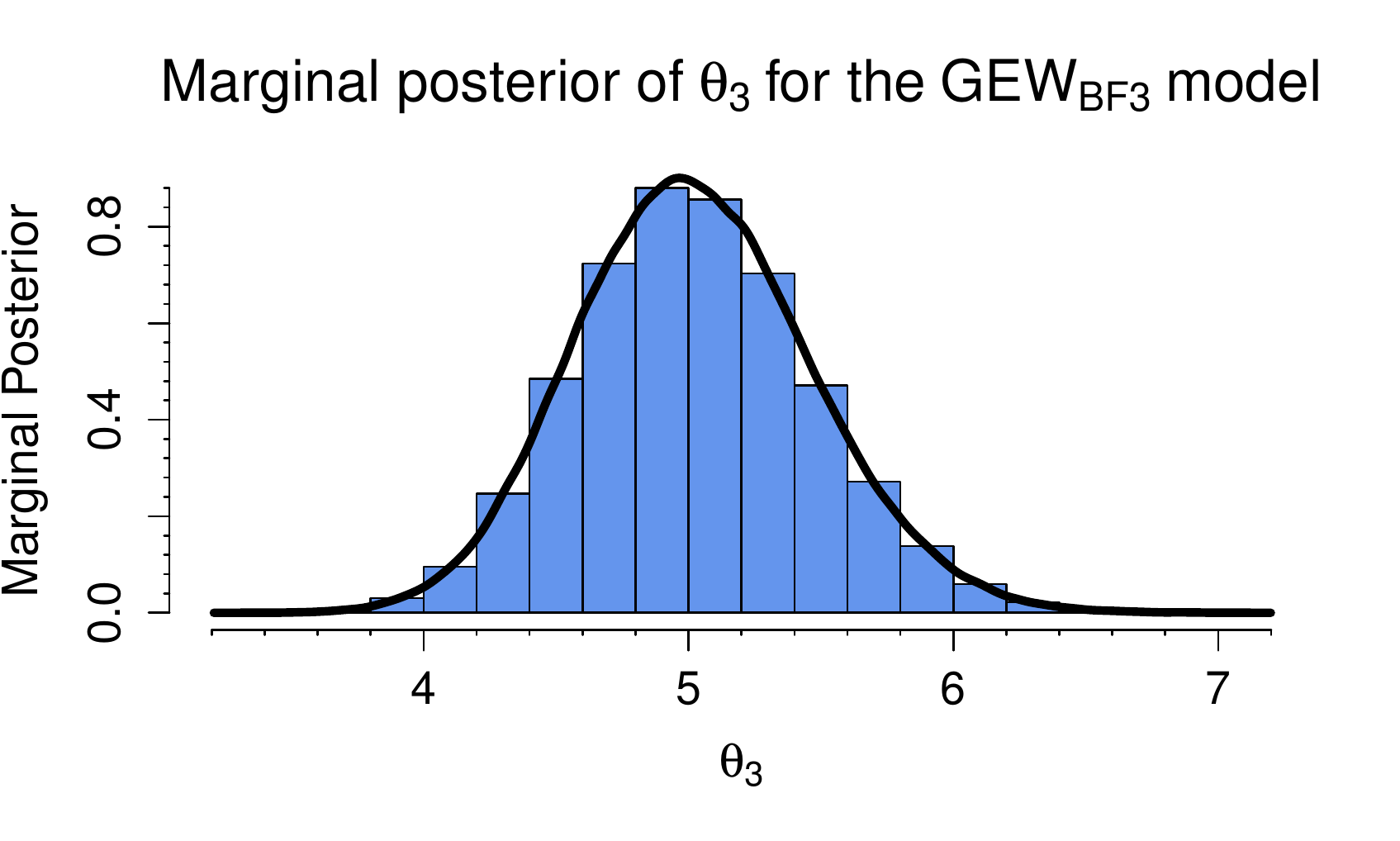}\includegraphics[scale=0.28]{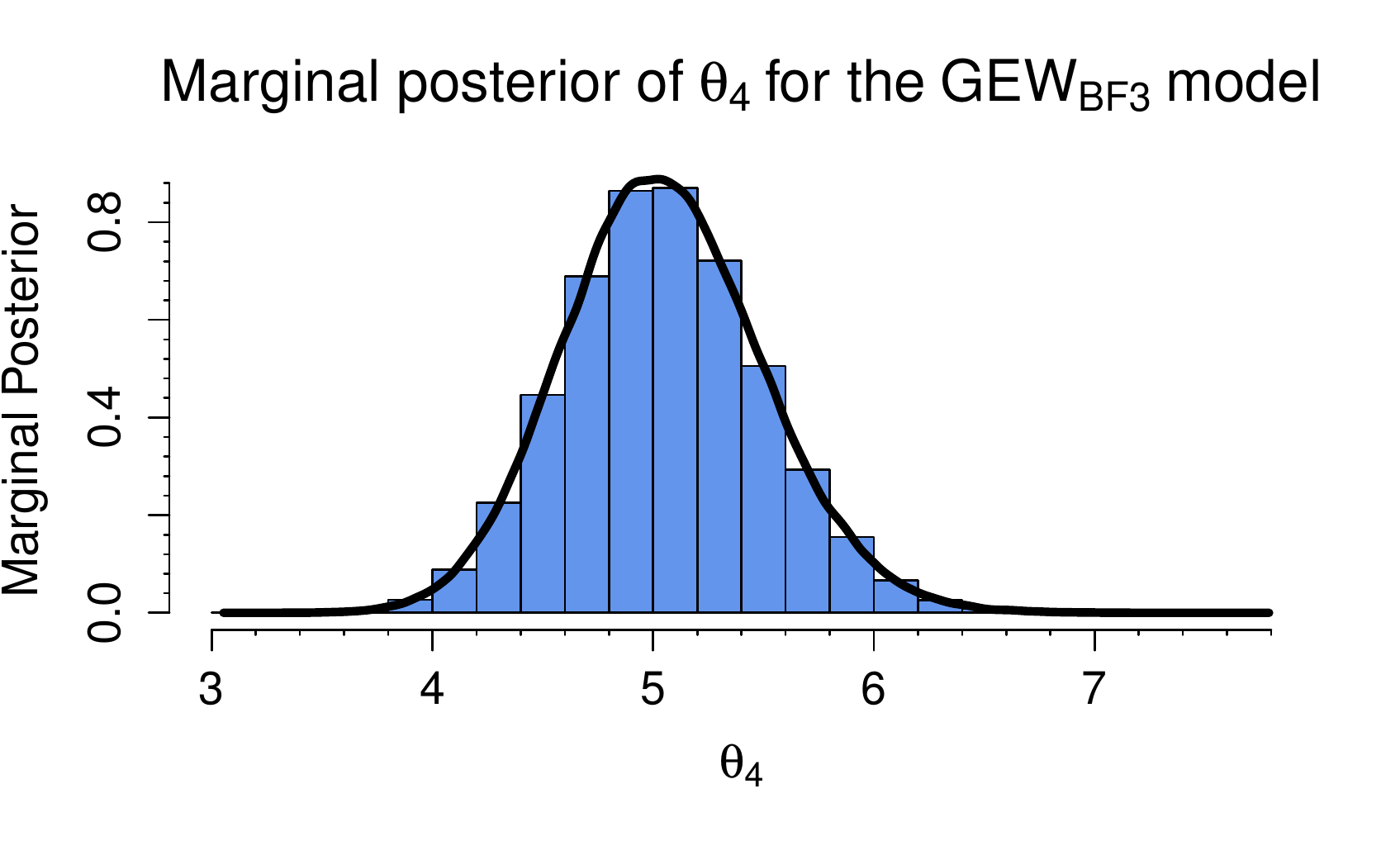}\includegraphics[scale=0.28]{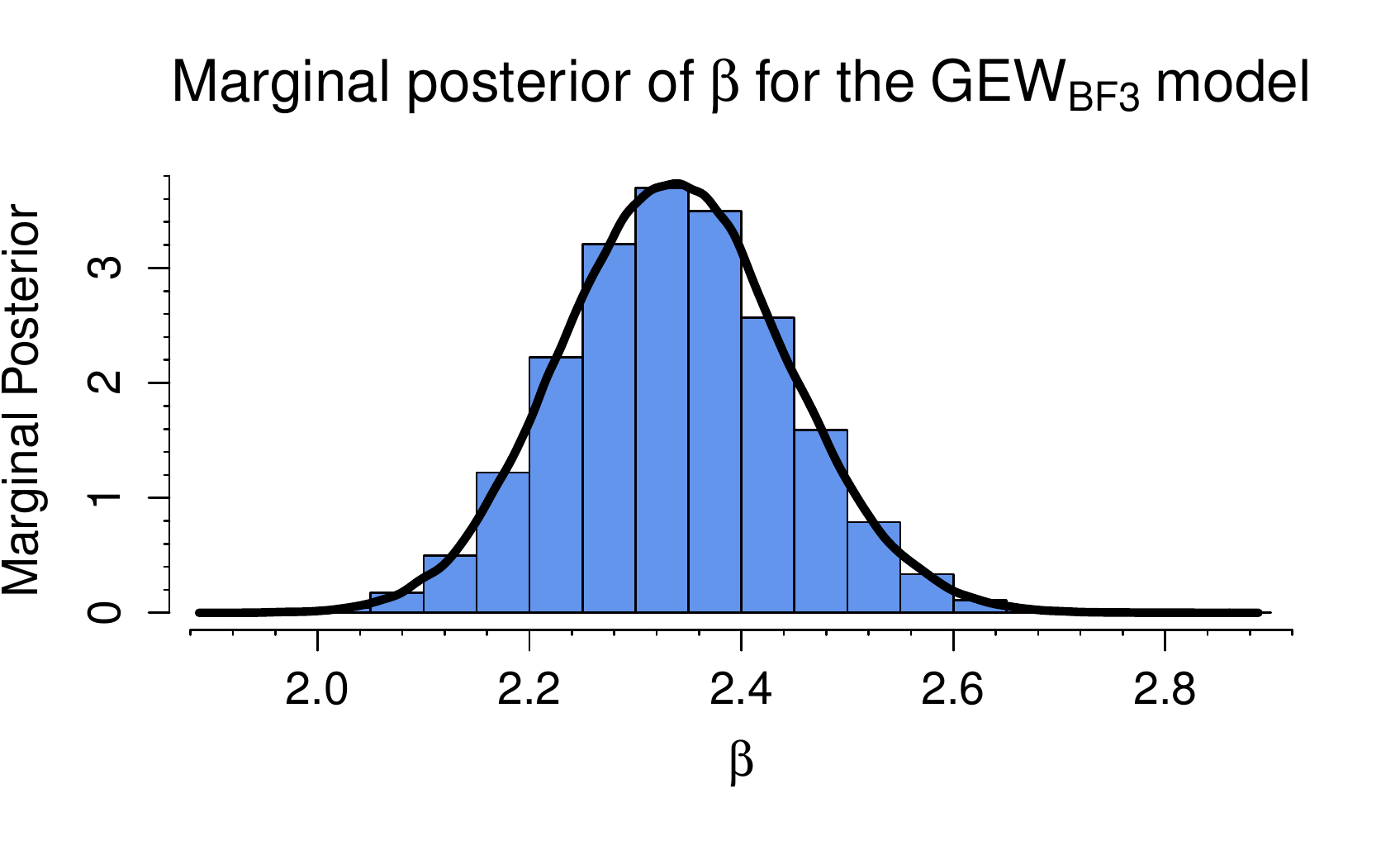}
\par\end{centering}
\caption{Marginal posterior distributions for the $\text{GEW}$ models.}
\label{Flo:GEWBF1_MARGINALS}
\end{sidewaysfigure}

\begin{sidewaysfigure}
\begin{centering}
\includegraphics[scale=0.28]{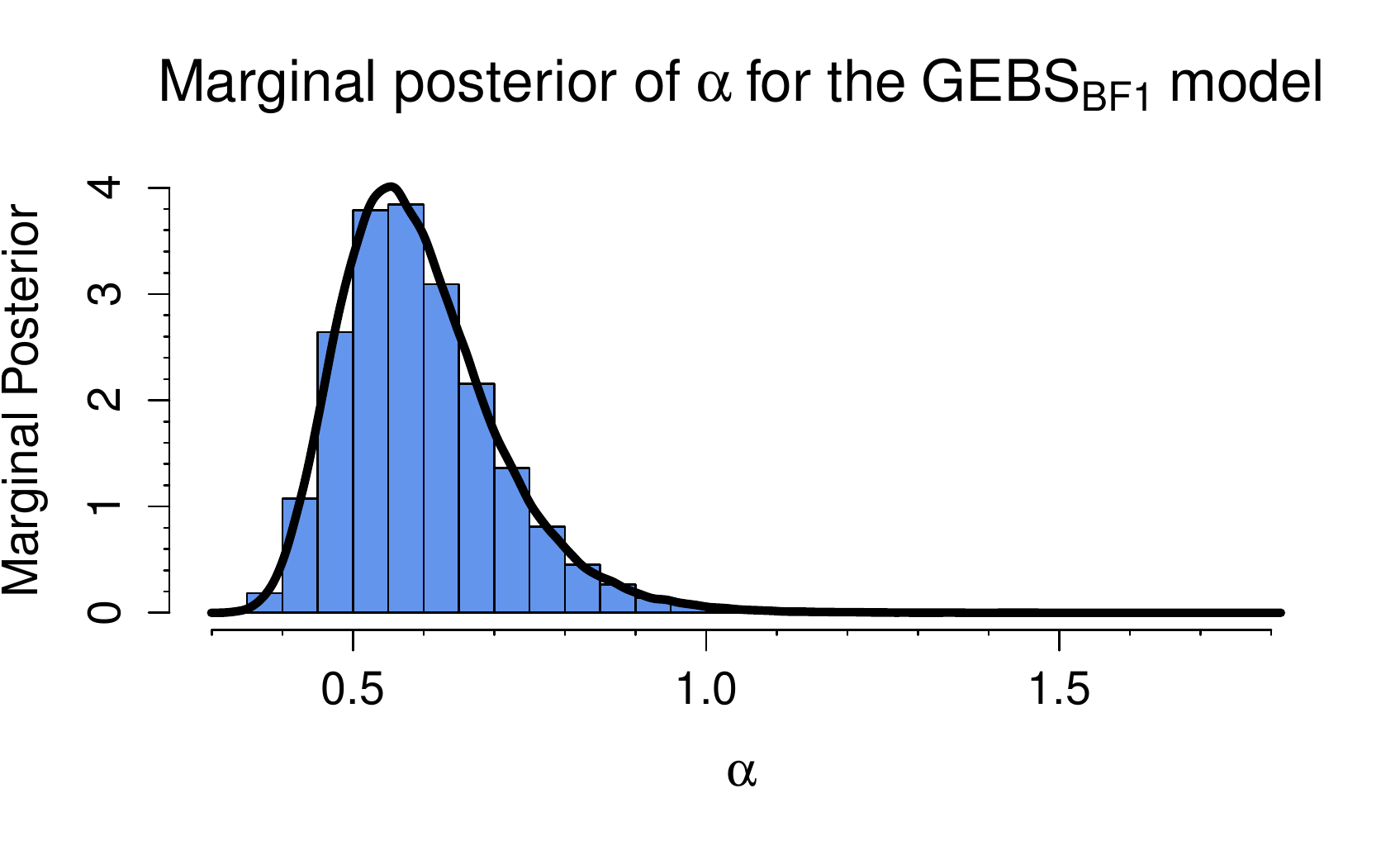}\includegraphics[scale=0.28]{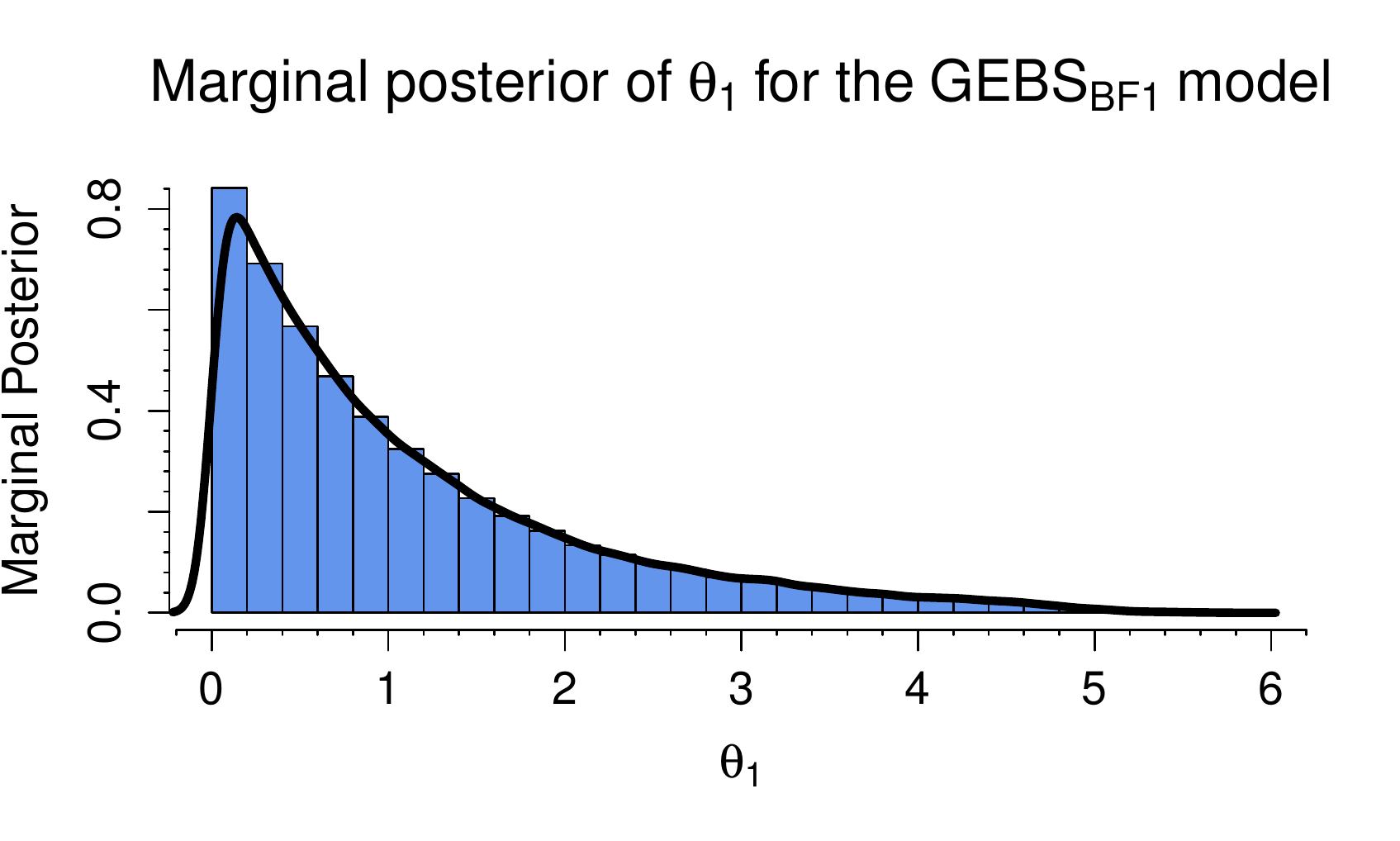}\includegraphics[scale=0.28]{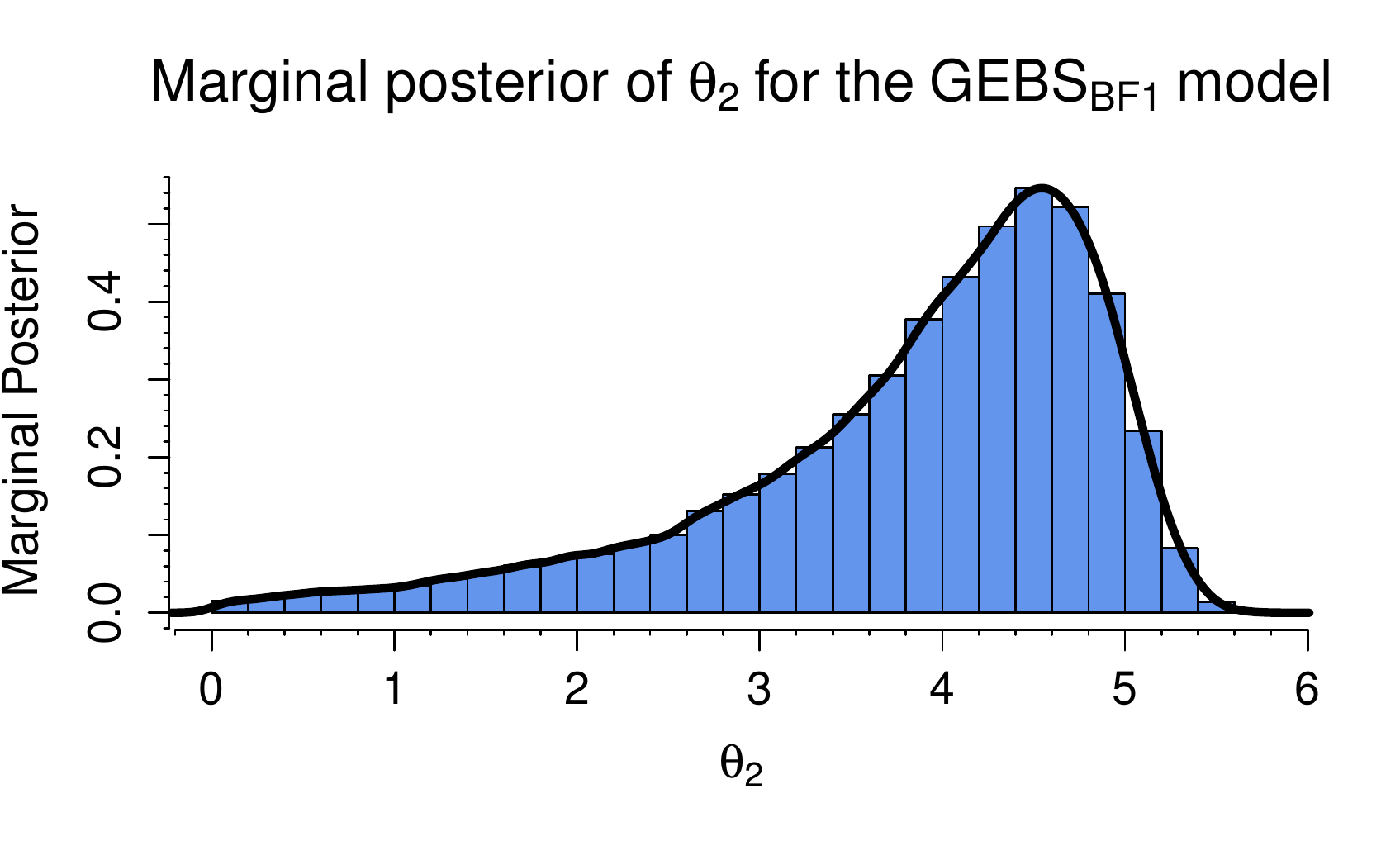}\includegraphics[scale=0.28]{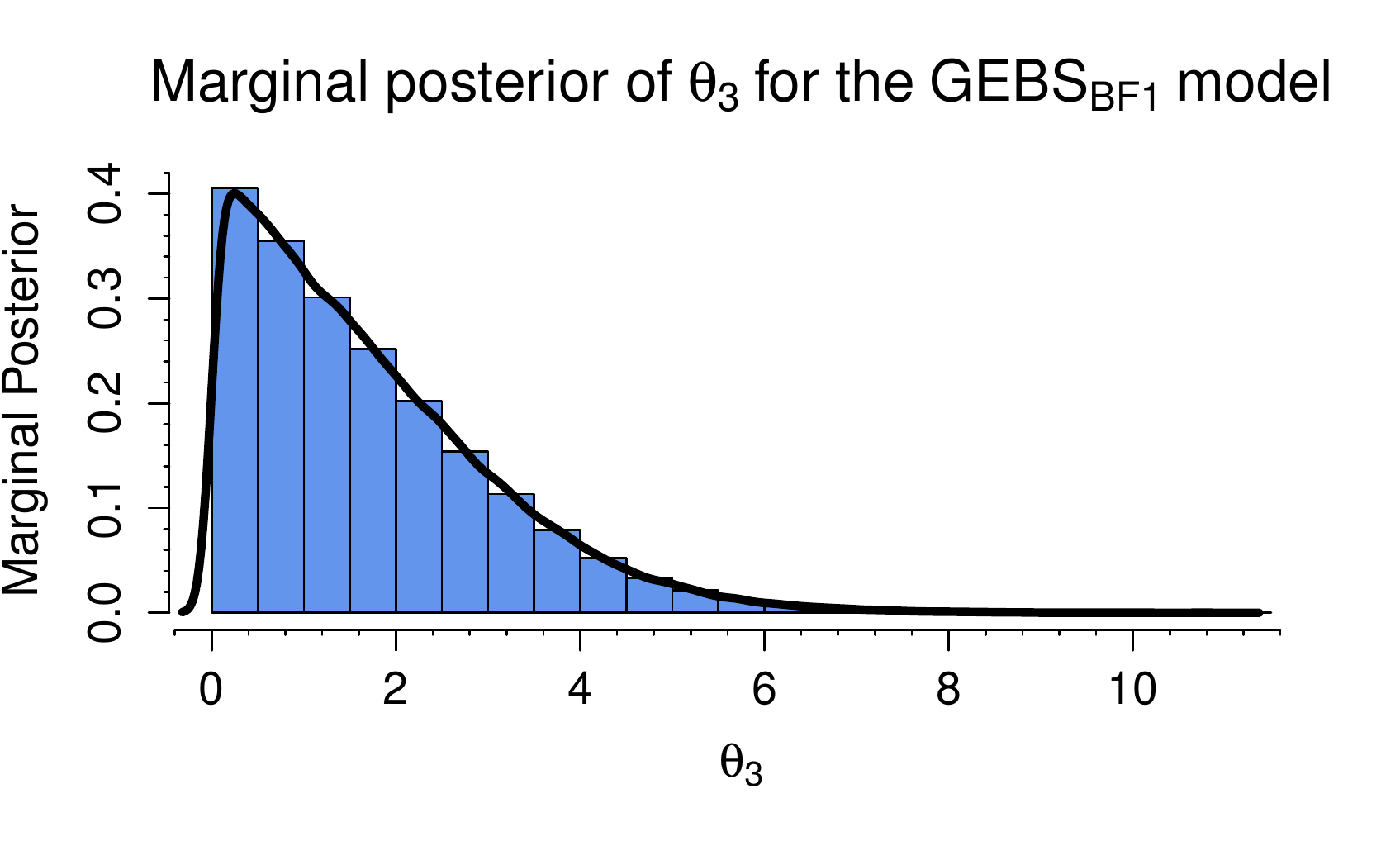}\includegraphics[scale=0.28]{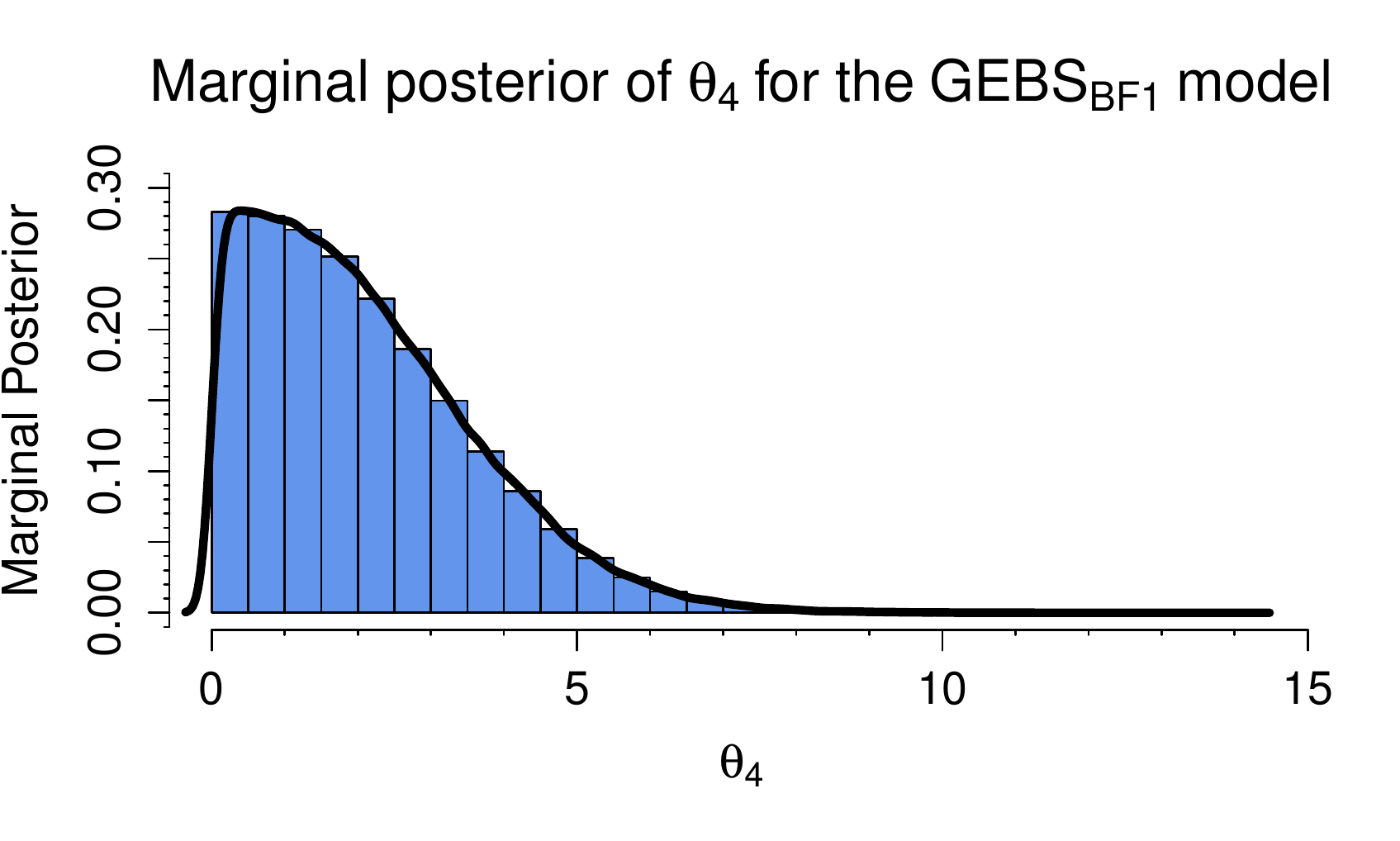}
\par\end{centering}
\begin{centering}
\includegraphics[scale=0.28]{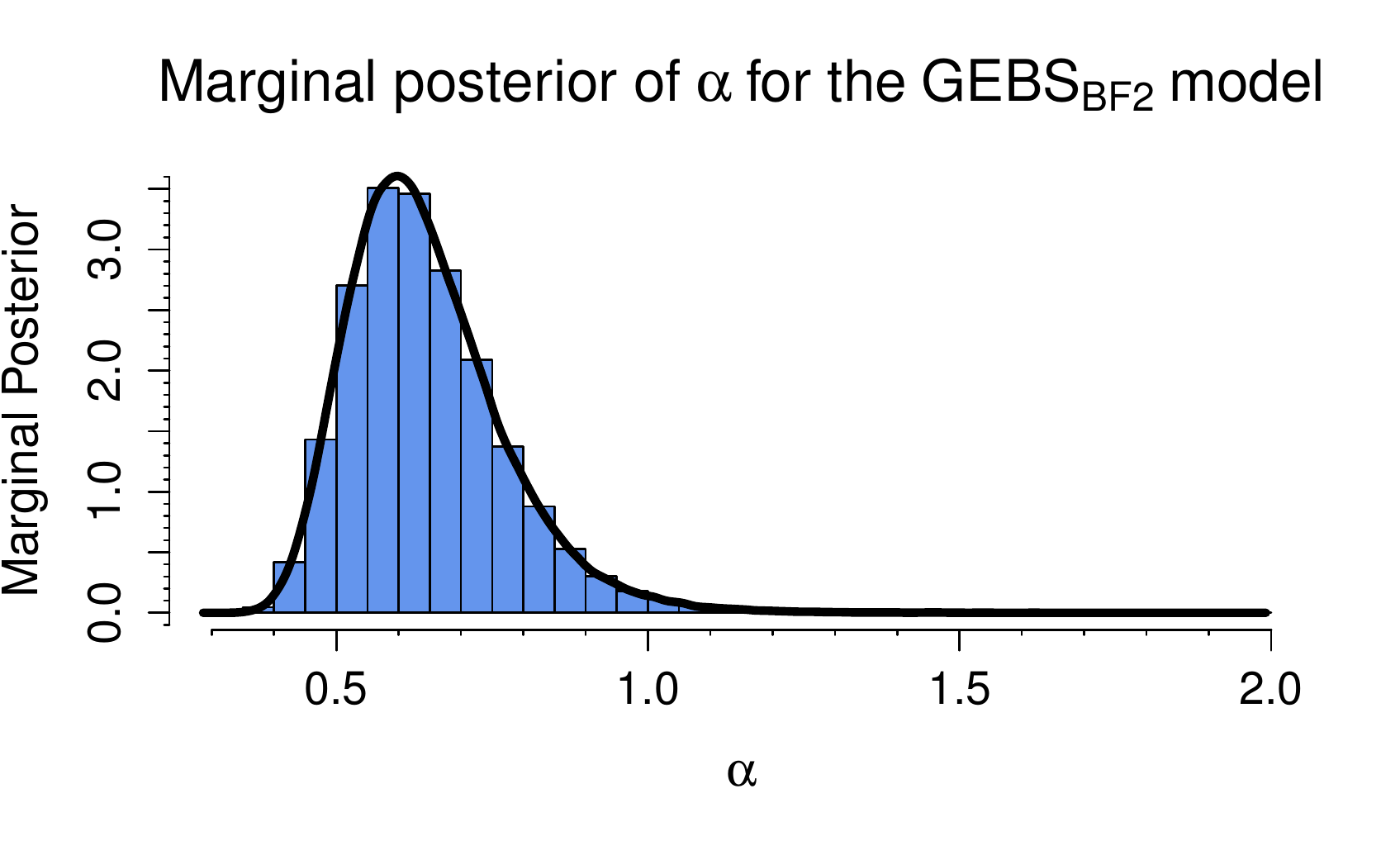}\includegraphics[scale=0.28]{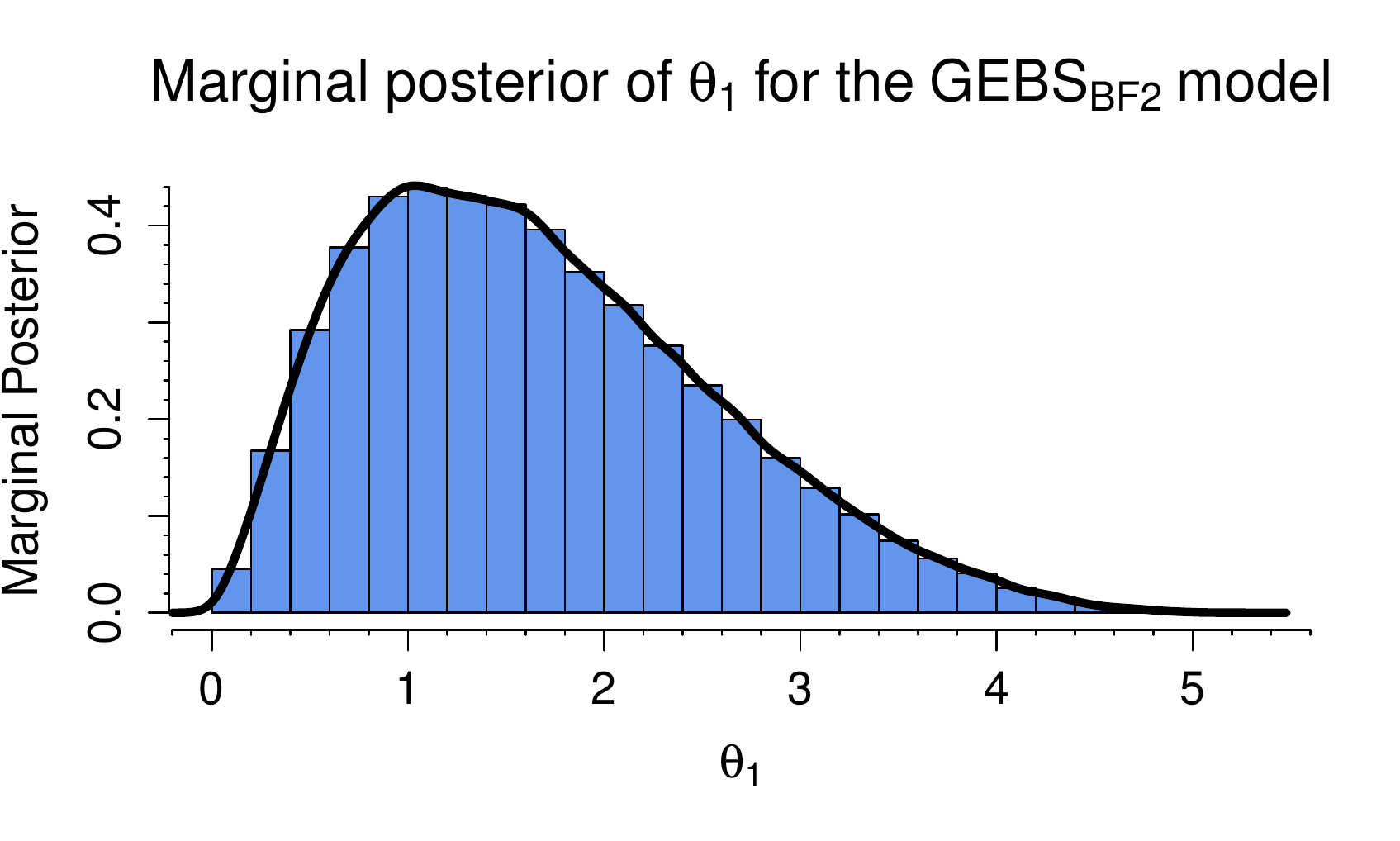}\includegraphics[scale=0.28]{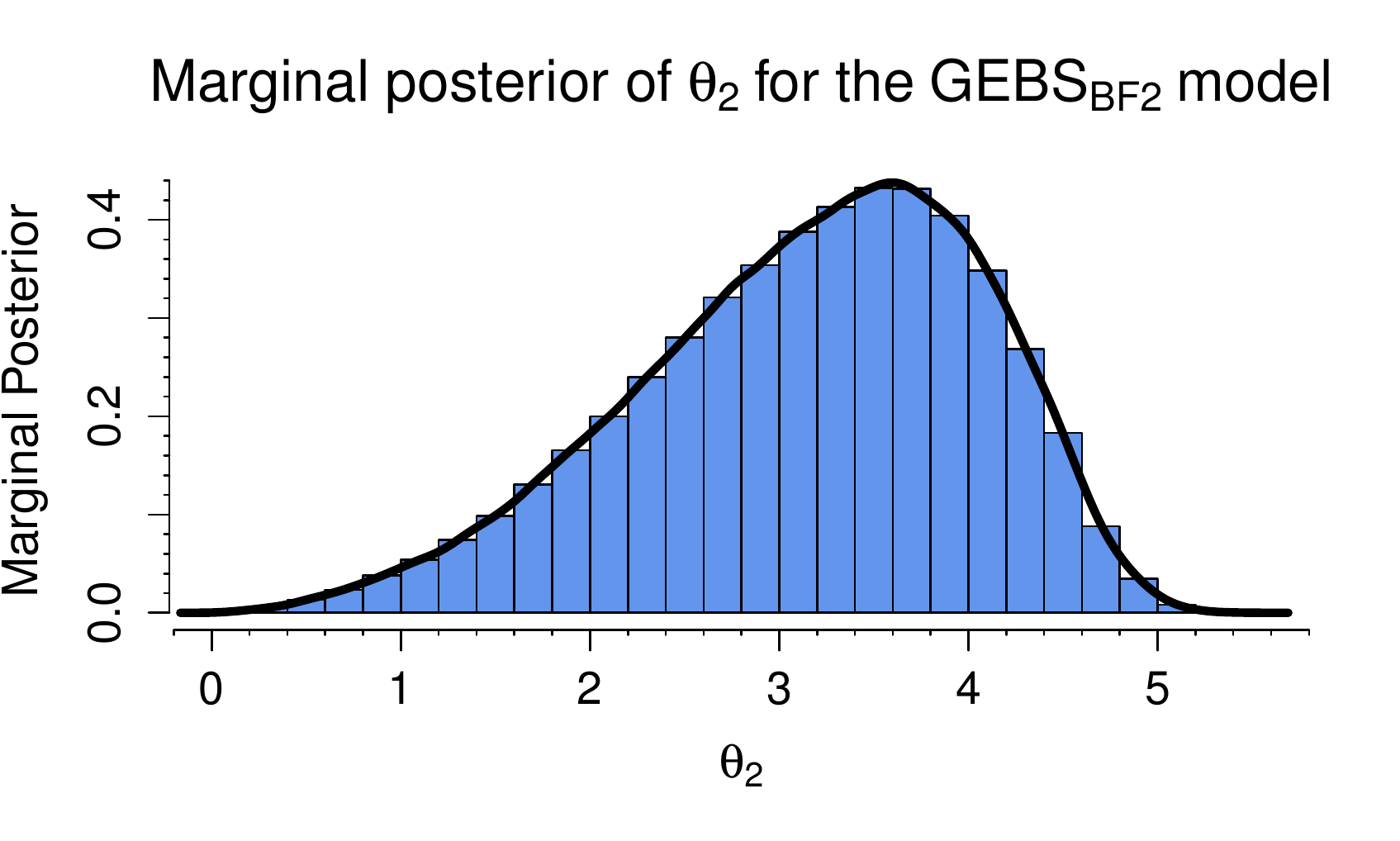}\includegraphics[scale=0.28]{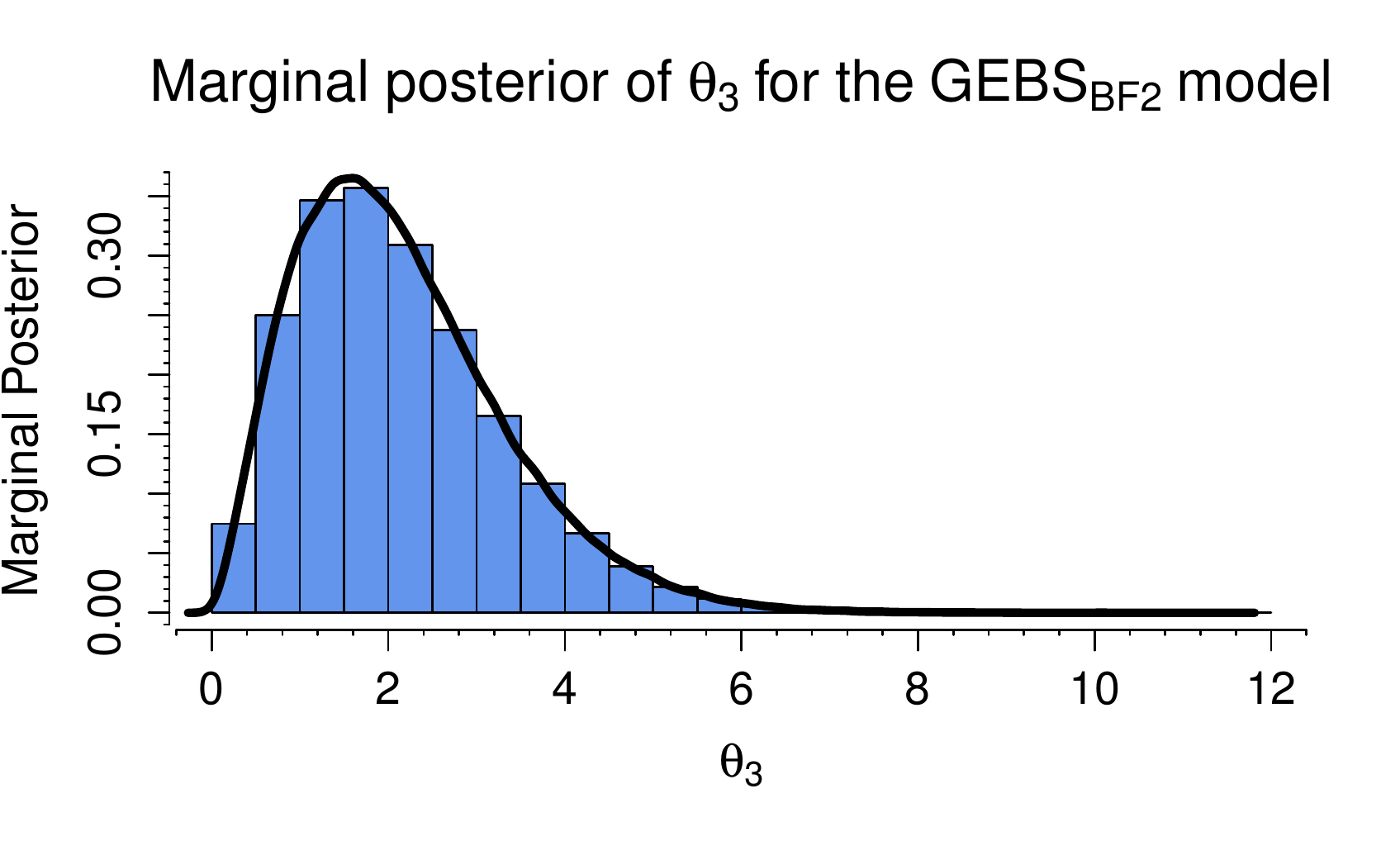}\includegraphics[scale=0.28]{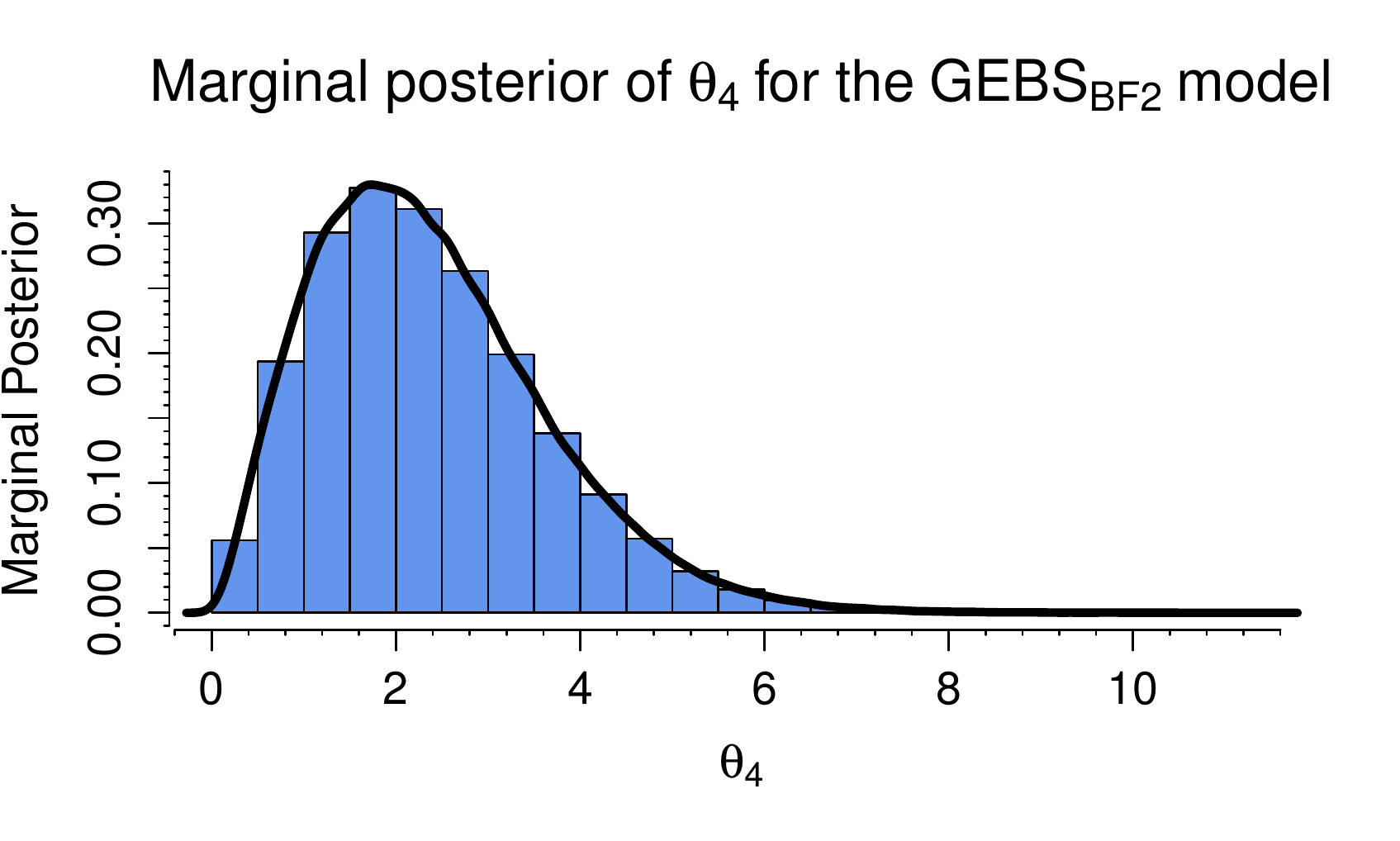}
\par\end{centering}
\begin{centering}
\includegraphics[scale=0.28]{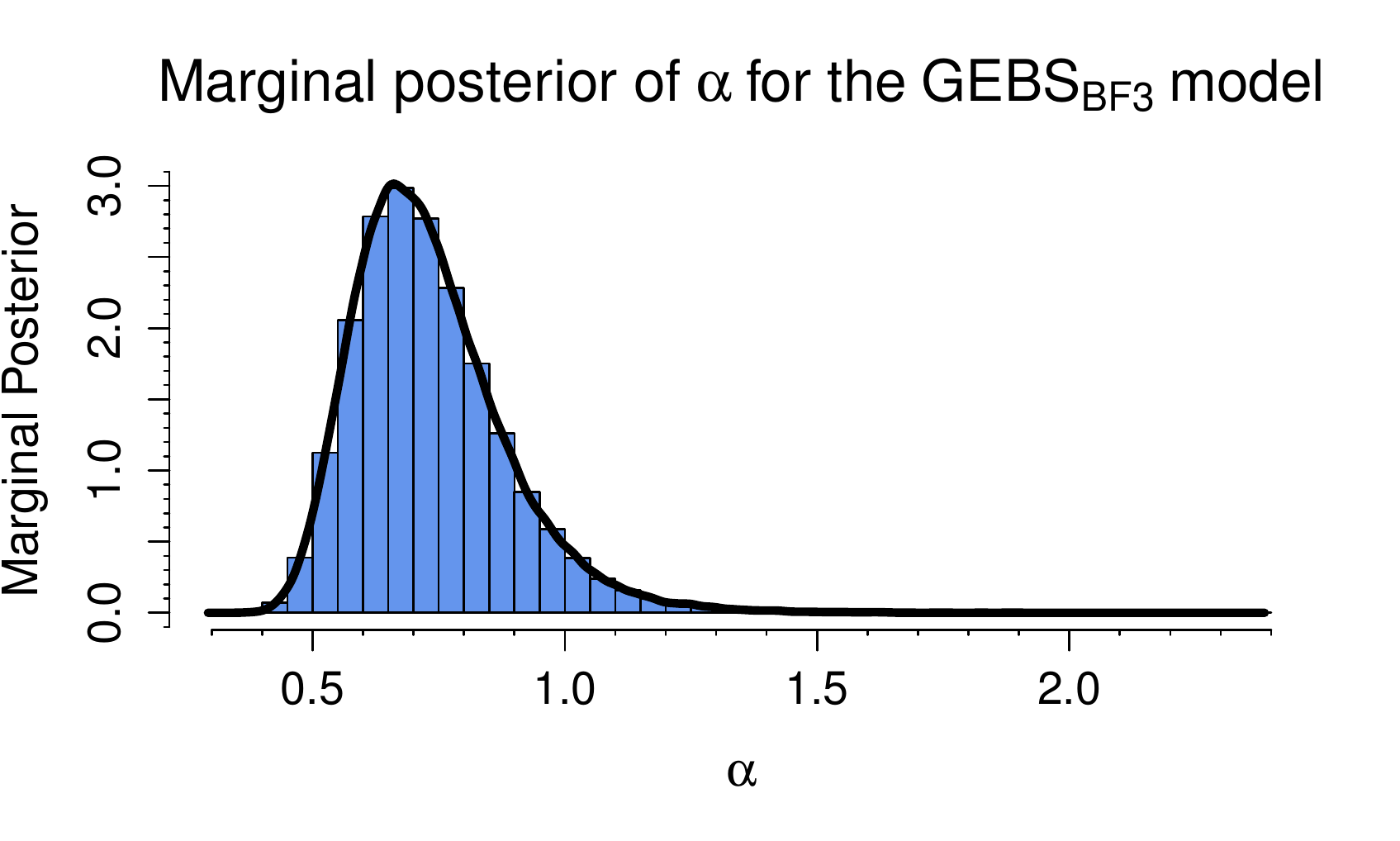}\includegraphics[scale=0.28]{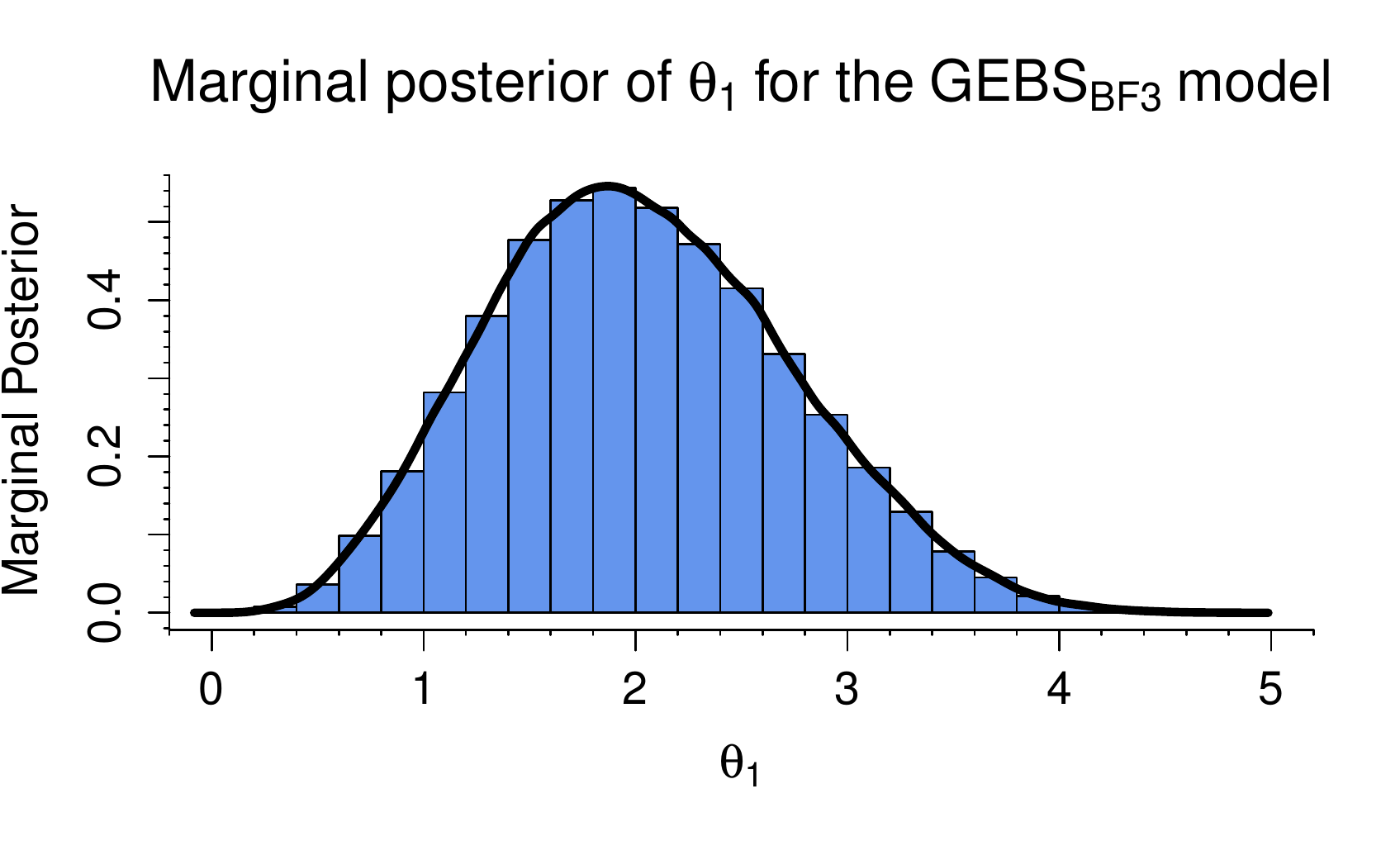}\includegraphics[scale=0.28]{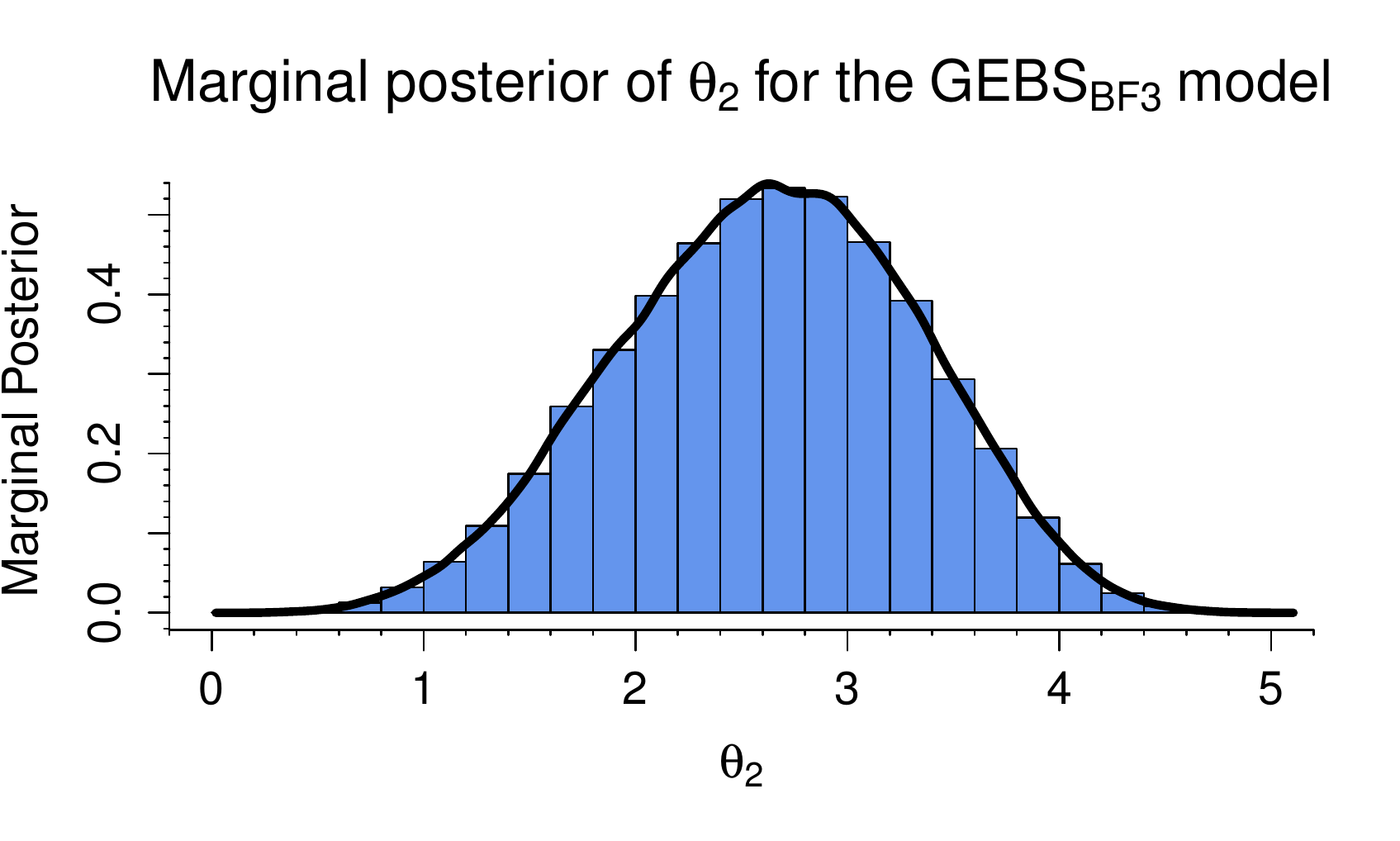}\includegraphics[scale=0.28]{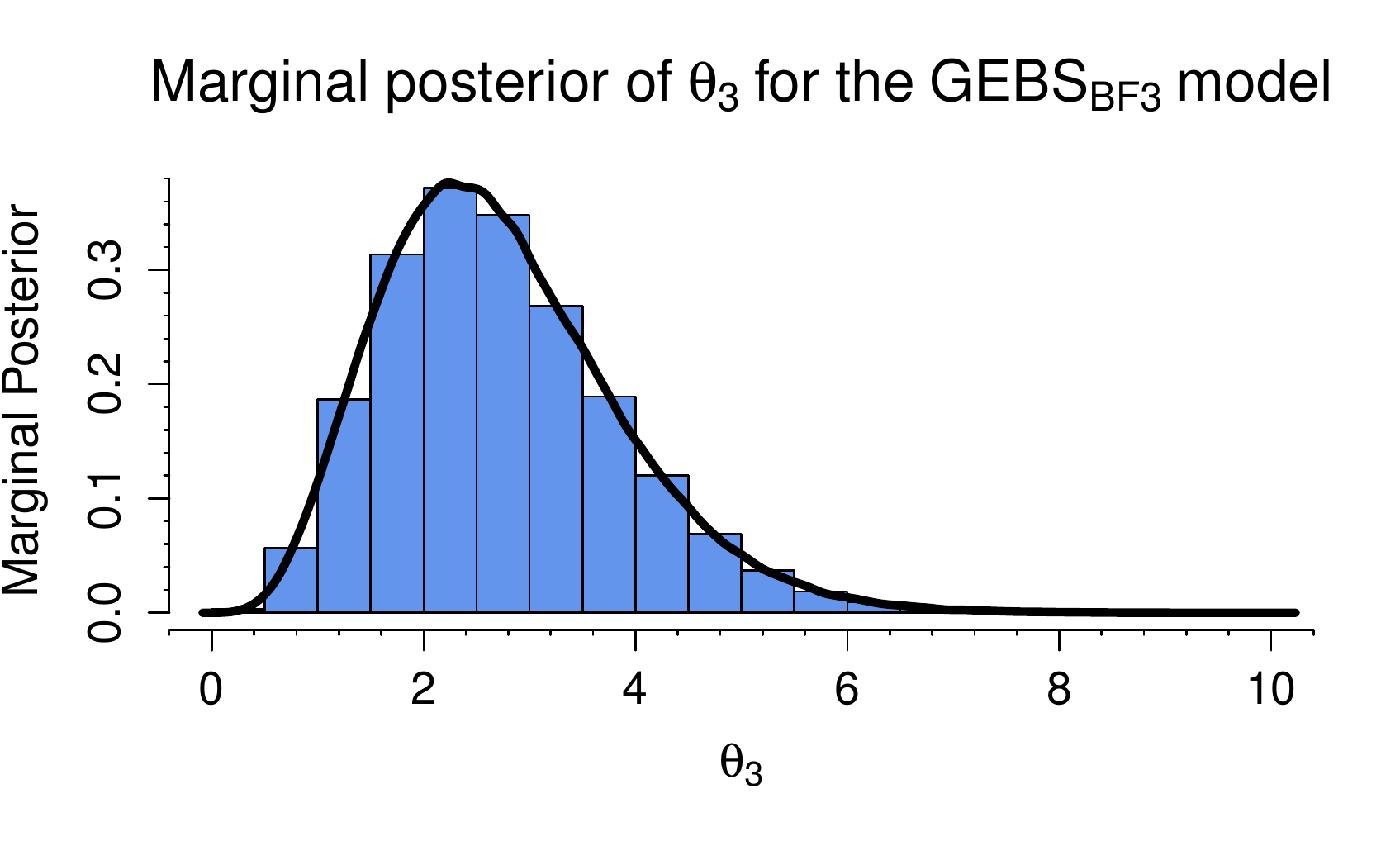}\includegraphics[scale=0.28]{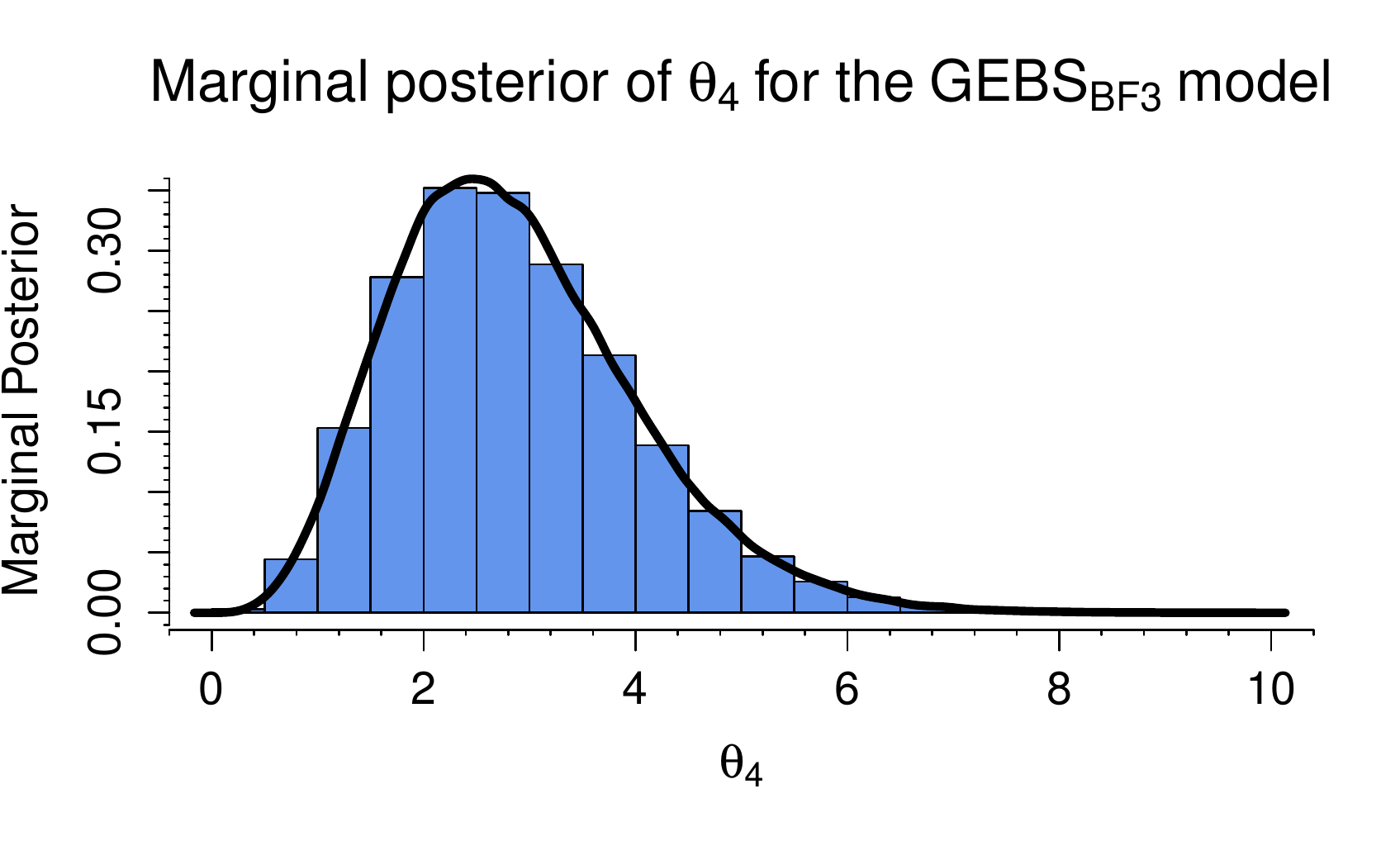}
\par\end{centering}
\caption{Marginal posterior distributions for the $\text{GEBS}$ models.}
 \label{Flo:GEBSBF3_MARGINALS}
\end{sidewaysfigure}

The marginal posterior distributions for the models are given in Figures
\ref{Flo:GEWBF1_MARGINALS} and \ref{Flo:GEBSBF3_MARGINALS}. It can
be noted that the models where flat priors are used produce fairly
skewed marginal posteriors. The marginal posteriors for the $\text{GEW}_{BF2}$,
$\text{GEBS}_{BF2}$ and $\text{GEBS}_{BF3}$ models are somewhat
less skewed, and those of the $\text{GEW}_{BF3}$ model are relatively
symmetric.

The DIC values for the models considered in this application are given
in Table \ref{Flo:BF_DIC}. The $\text{GEBS}_{BF1}$ model exhibits
the lowest DIC amongst the six models. It is clear that the DIC indicates
a strong preference towards the GEBS models in this case. Among the
GEW models, the $\text{GEW}_{BF1}$, exhibits the lowest DIC. Again,
it can be noted that the models with smaller variance subjective priors
have higher DIC values than the models with with flat priors or subjective
priors with a larger variance.

Considering all six models and keeping in mind that the $\text{GEBS}_{BF1}$
model has the lowest DIC, the guidelines in Burnham \& Anderson$^{5}$
indicate that $\text{GEBS}_{BF2}$ still has substantial support to
be used for inference. There is considerably less support to choose
$\text{GEBS}_{BF3}$, and the remaining three GEW models have almost
no support to be chosen. By investigating the GEW models separately,
we see that $\text{GEW}_{BF1}$ has the lowest DIC, there is fairly
substantial support for $\text{GEW}_{BF2}$, and considerably less
support for $\text{GEW}_{BF3}$.

Due to the large discrepancies between many of the prior distributions
in Table \ref{Flo:BF_PARMS} and the marginal posteriors produced
from the MCMC algorithm, the simple Monte Carlo estimator is omitted
from this application. Sampling from the prior for this estimator
causes computational issues for most of the models. However, for the
models where the marginal likelihood can be estimated, it is relatively
close to the approximation given by the Laplace-Metropolis estimator.

Table \ref{Flo:BF_LOG_MARGLIKE} shows the natural log of the marginal
likelihood estimates for the models, given by the Laplace-Metropolis
estimator and the harmonic mean estimator. The natural log of the
PPD estimates, which are required to calculate the posterior Bayes
factors, are also given. We number the models in this table in order
to simplify the notation for the Bayes factors to follow.

The Laplace-Metropolis estimator favours the $\text{GEW}_{BF2}$ model,
the harmonic mean estimator favours the $\text{GEW}_{BF1}$ model,
and the PPD estimate favours the $\text{GEBS}_{BF1}$ model. It is
interesting to note that the Laplace-Metropolis estimator does not
favour the models where flat priors are used. This can in part be
explained by the skewed marginal posterior distributions produced
by these models, seeing that the Laplace-Metropolis estimator works
well for symmetric distributions. For the harmonic mean estimator
and PPD estimate there is not such a distinctive preference towards
the GEBS models, as we have seen with the DIC. It must again be stressed
that the DIC and Bayes factors measure model fit differently, as is
explained in Spiegelhalter \textit{et al}.$^{2}$.

Table \ref{Flo:BF_LAPLACE} contains the Bayes factor values, given
by the Laplace-Metropolis estimator, harmonic mean estimator, and
the posterior Bayes factor values, for all combinations of the models
used in this application. Note that when using the Laplace-Metropolis
estimator there is positive to very strong evidence in favour of the
$\text{GEW}_{BF2}$ model. Considering the GEBS models seperately,
there is very strong evidence for the $\text{GEBS}_{BF2}$ model.
When using the harmonic mean estimator, we observe negligible to very
strong evidence for the $\text{GEW}_{BF1}$ model. Among only the
GEBS models, there is positive to strong evidence in favour of the
$\text{GEBS}_{BF1}$ model. It is clear that the Bayes factors, making
use of the harmonic mean estimator, also favour the models where flat
priors are used. This is in agreement with the conclusions from the
DIC. For the posterior Bayes factor values for the models, the PPD
estimates are used. Overall, the $\text{GEBS}_{BF1}$ model is supported
with negligible to strong evidence. For the GEW models separately,
we see positive to strong evidence for the $\text{GEW}_{BF1}$ model.
The posterior Bayes factors also favour the models where flat priors
are utilized. Table \ref{Flo:PMP_LAPLACE} contains the posterior
model probabilities (PMP) for each pair of models, given by the Laplace-Metropolis
estimator, the harmonic mean estimator, and when using the estimated
posterior Bayes factors.

\section{Conclusions\label{sec:BF_CONCLUSIONS}}

In this paper, the use of Bayes factors and the DIC for model selection
are compared in a Bayesian ALT setup. Two dual-stress models, namely
the GEW and GEBS models with gamma priors, are utilized for this comparison.
The posterior distributions for these models can not be written in
closed form, which complicates the calculation of the Bayes factors.
MCMC methods are employed to generate posterior samples to base inference
on. Methods for estimating the marginal likelihood, without further
complicating the sampling process, is explored. These methods include
a simple Monte Carlo estimator, the Laplace-Metropolis estimator,
the harmonic mean estimator, and a PPD estimate used for calculating
posterior Bayes factors. The models are applied to an ALT data set
where the stressors are temperature and relative humidity. Several
choices of hyperparameters are used in order to illustrate the use
of the DIC and Bayes factors in model selection. It is interesting
to note that the DIC shows definitive support for the GEBS models
above the GEW models. The models where flat priors are imposed on
the model parameters are favoured by the DIC. The different methods
for estimating the marginal likelihood give variable conclusions.
The simple Monte Carlo estimator is omitted in the application, due
to it causing computational issues for some of the models. The Laplace-Metropolis
estimator results in Bayes factors which show virtually no evidence
in favour of the models with flat priors. The Bayes factors produced
by the harmonic mean estimator and the posterior Bayes factors have
results that are more comparable to the DIC. Viewing the GEW and GEBS
models seperately, these Bayes factors also favour the models with
flat priors. The harmonic mean estimator shows more evidence in support
of the GEW models, whereas the posterior Bayes factors support the
GEBS models to a greater extent. It is interesting to note that the
conclusions from the posterior Bayes factors most closely relate to
those made by the DIC. 

\section*{References}
\begin{enumerate}
\item Kadane, J. B., \& Lazar, N. A. 2004. Methods and criteria for model
selection. \textit{Journal of the American Statistical Association},
\textbf{99}(465), 279--290.
\item Spiegelhalter, D. J., Best, N. G., Carlin, B. P., \& Van der Linde,
A. 2002. Bayesian measures of model complexity and fit. \textit{Journal
of the Royal Statistical Society, Series B}, \textbf{64}(4), 583--639.
\item Upadhyay, S. K., \& Mukherjee, B. 2010. Bayes analysis and comparison
of accelerated Weibull and accelerated Birnbaum-Saunders models. \textit{Communications
in Statistics - Theory and Methods}, \textbf{39}, 195--213.
\item Bernardo, J. M., \& Smith, A. F. M. 1994. \textit{Bayesian Theory}.
New York: Wiley.
\item Burnham, K. P., \& Anderson, D. R. 1998. \textit{Model Selection and
Inference}. New York: Springer.
\item Akaike, H. 1973. Information theory and an extension of the maximum
likelihood principle. Pages 267--281 of: Petrov, B. N., \& Csáki,
F. (eds), \textit{Proceedings of the 2nd International Symposium on
Information Theory}.
\item Barriga, G. D. C., Ho, L. L., \& Cancho, V. G. 2008. Planning accelerated
life tests under exponentiated- Weibull-Arrhenius model. \textit{International
Journal of Quality \& Reliability Management}, \textbf{25}(6), 636--653.
\item Soyer, R., Erkanli, A., \& Merrick, J. R. 2008. Accelerated life tests:
Bayesian models. \textit{Encyclopedia of Statistics in Quality and
Reliability}, \textbf{1}, 20--30.
\item Jeffreys, H. 1961. \textit{Theory of Probability}. Oxford: Oxford
University Press.
\item Kass, R. E., \& Raftery, A. E. 1995. Bayes factors. \textit{Journal
of the American Statistical Association}, \textbf{90}(430), 773--795.
\item Ntzoufras. 2009. \textit{Bayesian Modeling Using WinBUGS}. Hoboken:
Wiley.
\item DeGroot, M. H. 1970. \textit{Optimal Statistical Decisions}. New York:
McGraw-Hill.
\item Zellner, A. 1971. \textit{An Introduction to Bayesian Inference in
Econometrics}. New York: John Wiley.
\item Geisser, S., \& Eddy, W. F. 1979. A predictive approach to model selection.
\textit{Journal of the American Statistical Association}, \textbf{74}(365),
153--160.
\item Aitkin, M. 1991. Posterior Bayes factors. \textit{Journal of the Royal
Statistical Society, Series B}, \textbf{53}(1), 111--142.
\item O\textquoteright Hagan, A. 1995. Fractional Bayes factors for model
comparison. \textit{Journal of the Royal Statistical Society, Series
B}, \textbf{57}(1), 99--138.
\item Berger, J. O., \& Pericchi, L. R. 1996. The intrinsic Bayes factor
for model selection and prediction. \textit{Journal of the American
Statistical Association}, \textbf{91}(433), 109--122.
\item Smit, N. 2021. \textit{Accelerated life testing using the Eyring model
for the Weibull and Birnbaum-Saunders distributions}. Ph.D. thesis,
North-West University.
\item Smit, N., \& Raubenheimer, L. 2021. Bayesian accelerated life testing:
A generalized Eyring-Birnbaum- Saunders model. \textit{Quality and
Reliability Engineering International}. DOI: https://doi.org/10.1002/qre.2970
\item Owen, W. J., \& Padgett, W. J. 2000. A Birnbaum-Saunders accelerated
life model. \textit{IEEE Transactions on Reliability}, \textbf{49}(2),
224--229.
\item Sun, T., \& Shi, Y. 2016. Estimation for Birnbaum-Saunders distribution
in simple step stress-accelerated life test with type-II censoring.
\textit{Communications in Statistics - Simulation and Computation},
\textbf{45}, 880-- 901.
\item Sha, N. 2018. Statistical inference for progressive stress accelerated
life testing with Birnbaum-Saunders distribution. \textit{Stats},
\textbf{1}, 189--203.
\item Escobar, L. A., \& Meeker, W. Q. 2006. A review of accelerated test
models. \textit{Statistical Science}, \textbf{21}(4), 552--577.
\item Mazzuchi, T. A., Soyer, R., \& Vopatek, A. L. 1997. Linear Bayesian
inference for accelerated Weibull model. \textit{Lifetime Data Analysis},
\textbf{3}, 63--75.
\item Raftery, A. E. 1996. Hypothesis testing and model selection. Pages
163--188 of: Gilks, W. R., Richardson, S., \& Spiegelhalter, D. J.
(eds), \textit{Markov Chain Monte Carlo in Practice}. Suffolk: Chapman
\& Hall.
\item Lewis, S. M., \& Raftery, A. E. 1997. Estimating Bayes factors via
posterior simulation with the Laplace- Metropolis estimator. \textit{Journal
of the American Statistical Association},\textbf{ 92}(438), 648--655.
\item Newton, M. A., \& Raftery, A. E. 1994. Approximate Bayesian inference
with the weighted likelihood bootstrap. \textit{Journal of the Royal
Statistical Society, Series B}, \textbf{56}(1), 3--48.
\item Raftery, A. E., Newton, M. A., Satagopan, J. M., \& Krivitsky, P.
N. 2007. Estimating the integrated likelihood via posterior simulation
using the harmonic mean identity. \textit{Bayesian Statistics}, \textbf{8},
1--45.
\item Gelfand, A. E., \& Dey, D. K. 1994. Bayesian model choice: Asymptotics
and exact calculations. \textit{Journal of the Royal Statistical Society,
Series B}, \textbf{56}, 501--514.
\item Weinberg, M. D. 2012. Computing the Bayes Factor from a Markov chain
Monte Carlo simulation of the posterior distribution. \textit{Bayesian
Analysis}, \textbf{7}(3), 737--770.
\item ReliaSoft. 2020. Accelerated testing data analysis without a known
physical failure model. (Accessed October 15, 2020). https://www.reliasoft.com/resources/resource-center/accelerated-testing-dataanalysis-without-a-known-physical-failure-model
\item Brooks, S. P., \& Gelman, A. 1998. General methods for monitoring
convergence of iterative simulations. \textit{Journal of Computational
and Graphical Statistics}, \textbf{7}(4), 434--455.
\end{enumerate}
\newpage{}
\begin{table}[H]
\caption{Failure times for some device.}
\label{Flo:BF_DATA}\medskip{}

\centering{}%
\begin{tabular}{|l|c|c|c|c|c|c|c|}
\hline 
{\small{}\#} & {\small{}Temperature (K)} & {\small{}Humidity} & {\small{}Failure time} & {\small{}\#} & {\small{}Temperature (K)} & {\small{}Humidity} & {\small{}Failure time}\tabularnewline
\hline 
1 & 333 & 0.9 & 521 & 12 & 353 & 0.8 & 504\tabularnewline
2 & 333 & 0.9 & 561 & 13 & 353 & 0.9 & 115\tabularnewline
3 & 333 & 0.9 & 575 & 14 & 353 & 0.9 & 119\tabularnewline
4 & 333 & 0.9 & 599 & 15 & 353 & 0.9 & 150\tabularnewline
5 & 333 & 0.9 & 609 & 16 & 353 & 0.9 & 152\tabularnewline
6 & 333 & 0.9 & 684 & 17 & 353 & 0.9 & 153\tabularnewline
7 & 333 & 0.9 & 709 & 18 & 353 & 0.9 & 155\tabularnewline
8 & 333 & 0.9 & 713 & 19 & 353 & 0.9 & 156\tabularnewline
9 & 353 & 0.8 & 345 & 20 & 353 & 0.9 & 164\tabularnewline
10 & 353 & 0.8 & 357 & 21 & 353 & 0.9 & 199\tabularnewline
11 & 353 & 0.8 & 439 &  &  &  & \tabularnewline
\hline 
\end{tabular}
\end{table}

\newpage{}
\begin{table}[H]
\caption{Prior specifications for the Bayes factors application.}
\textbf{\label{Flo:BF_PARMS}}
\centering{}\medskip{}
\begin{tabular}{|c|c|c|c|c|c|}
\hline 
Model & $\theta_{1}$ & $\theta_{2}$ & $\theta_{3}$ & \textbf{$\theta_{4}$} & \textbf{$\beta$}\tabularnewline
\hline 
$\text{GEW}_{BF1}$ & $\varGamma(1,0.001)$ & $\varGamma(1,0.001)$ & $\varGamma(1,0.001)$ & $\varGamma(1,0.001)$ & $\varGamma(1,0.001)$\tabularnewline
$\text{GEW}_{BF2}$ & $\varGamma(5,1)$ & $\varGamma(5,1)$ & $\varGamma(5,1)$ & $\varGamma(5,1)$ & $\varGamma(5,1)$\tabularnewline
$\text{GEW}_{BF3}$ & $\varGamma(125,25)$ & $\varGamma(125,25)$ & $\varGamma(125,25)$ & $\varGamma(125,25)$ & $\varGamma(125,25)$\tabularnewline
$\text{GEBS}_{BF1}$ & $\varGamma(1,0.001)$ & $\varGamma(1,0.001)$ & $\varGamma(1,0.001)$ & $\varGamma(1,0.001)$ & $\varGamma(1,0.001)$\tabularnewline
$\text{GEBS}_{BF2}$ & $\varGamma(2.5,0.5)$ & $\varGamma(2.5,0.5)$ & $\varGamma(2.5,0.5)$ & $\varGamma(2.5,0.5)$ & $\varGamma(2.5,0.5)$\tabularnewline
$\text{GEBS}_{BF3}$ & $\varGamma(5,1)$ & $\varGamma(5,1)$ & $\varGamma(5,1)$ & $\varGamma(5,1)$ & $\varGamma(5,1)$\tabularnewline
\hline 
\end{tabular}
\end{table}
\newpage{}
\newpage{}
\newpage{}
\begin{table}[H]
\caption{Deviance information criterion for the Bayes factors application.}
\textbf{\label{Flo:BF_DIC}}
\begin{centering}
\medskip{}
\par\end{centering}
\centering{}%
\begin{tabular}{|c|c|c|c|c|c|c|}
\hline 
\textbf{Model} & $\text{GEW}_{BF1}$ & $\text{GEW}_{BF2}$ & $\text{GEW}_{BF3}$ & $\text{GEBS}_{BF1}$ & $\text{GEBS}_{BF2}$ & $\text{GEBS}_{BF3}$\tabularnewline
\hline 
\textbf{DIC} & 278.4 & 280.9 & 284.6 & 248.9 & 250.8 & 253.3\tabularnewline
\hline 
\end{tabular}
\end{table}
\newpage{}
\begin{table}[H]
\caption{Natural log of the marginal likelihood estimates for the models.}
\label{Flo:BF_LOG_MARGLIKE}\medskip{}

\centering{}%
\begin{tabular}{|c|c|>{\centering}p{3.7cm}|>{\centering}p{3.7cm}|>{\centering}p{3.7cm}|}
\hline 
\textbf{Model \#} & \textbf{Model} & \textbf{\small{}Laplace-Metropolis estimator} & \textbf{Harmonic mean estimator} & \textbf{PPD estimate}\tabularnewline
\hline 
\multirow{1}{*}{{\small{}1}} & $\text{GEW}_{BF1}$ & -166.7920 & -141.5369 & -137.1797\tabularnewline
{\small{}2} & $\text{GEW}_{BF2}$ & -145.1006 & -141.6805 & -138.7164\tabularnewline
{\small{}3} & $\text{GEW}_{BF3}$ & -176.4858 & -145.1727 & -141.3472\tabularnewline
{\small{}4} & $\text{GEBS}_{BF1}$ & -170.1288 & -142.2716 & -136.8164\tabularnewline
{\small{}5} & $\text{GEBS}_{BF2}$ & -147.7771 & -144.2880 & -137.6861\tabularnewline
{\small{}6} & $\text{GEBS}_{BF3}$ & -151.8292 & -146.6435 & -138.8855\tabularnewline
\hline 
\end{tabular}
\end{table}
\newpage{}
\begin{sidewaystable}[H]
\caption{Bayes factors.}
\label{Flo:BF_LAPLACE}\medskip{}

\centering{}{\small{}}%
\begin{tabular}{|c|c|c|c|c|c|c|}
\hline 
 & \multicolumn{2}{c|}{\textbf{\small{}Laplace-Metropolis estimates}} & \multicolumn{2}{c|}{\textbf{\small{}Harmonic mean estimates}} & \multicolumn{2}{c|}{\textbf{\small{}PPD estimates}}\tabularnewline
\hline 
 & {\small{}$B_{ij}$} & {\small{}Interpretation} & {\small{}$B_{ij}$} & {\small{}Interpretation} & {\small{}$B_{ij}$} & {\small{}Interpretation}\tabularnewline
\hline 
\multirow{1}{*}{{\small{}$B_{12}$}} & {\small{}$3.7979\times10^{-10}$} & {\small{}Very strong evidence for model 2} & {\small{}1.1543} & {\small{}Negligible evidence for model 1} & {\small{}4.6489} & {\small{}Positive evidence for model 1}\tabularnewline
{\small{}$B_{13}$} & {\small{}$1.6217\times10^{4}$} & {\small{}Very strong evidence for model $1$} & {\small{}37.9317} & {\small{}Strong evidence for model 1} & {\small{}64.5548} & {\small{}Strong evidence for model 1}\tabularnewline
{\small{}$B_{14}$} & {\small{}28.1309} & {\small{}Strong evidence for model 1} & {\small{}2.0847} & {\small{}Negligible evidence for model 1} & {\small{}0.6954} & {\small{}Negligible evidence for model 4}\tabularnewline
{\small{}$B_{15}$} & {\small{}$5.5201\times10^{-9}$} & {\small{}Very strong evidence for model 5} & {\small{}15.6587} & {\small{}Positive evidence for model 1} & {\small{}1.6592} & {\small{}Negligible evidence for model 1}\tabularnewline
{\small{}$B_{16}$} & {\small{}$3.1751\times10^{-7}$} & {\small{}Very strong evidence for model 6} & {\small{}165.1082} & {\small{}Very strong evidence for model 1} & {\small{}5.5059} & {\small{}Positive evidence for model 1}\tabularnewline
{\small{}$B_{23}$} & {\small{}$4.2700\times10^{13}$} & {\small{}Very strong evidence for model 2} & {\small{}32.8600} & {\small{}Strong evidence for model 2} & {\small{}13.8860} & {\small{}Positive evidence for model 2}\tabularnewline
{\small{}$B_{24}$} & {\small{}$7.4069\times10^{10}$} & {\small{}Very strong evidence for model 2} & {\small{}1.8060} & {\small{}Negligible evidence for model 2} & {\small{}0.1496} & {\small{}Positive evidence for model 4}\tabularnewline
{\small{}$B_{25}$} & {\small{}14.5346} & {\small{}Positive evidence for model 2} & {\small{}13.5651} & {\small{}Positive evidence for model 2} & {\small{}0.3569} & {\small{}Negligible evidence for model 5}\tabularnewline
{\small{}$B_{26}$} & {\small{}836.0159} & {\small{}Very strong evidence for model 2} & {\small{}143.0325} & {\small{}Strong evidence for model 2} & {\small{}1.1843} & {\small{}Negligible evidence for model 2}\tabularnewline
{\small{}$B_{34}$} & {\small{}0.0017} & {\small{}Very strong evidence for model 4} & {\small{}0.0550} & {\small{}Positive evidence for model 4} & {\small{}0.0108} & {\small{}Strong evidence for model 4}\tabularnewline
{\small{}$B_{35}$} & {\small{}$3.4039\times10^{-13}$} & {\small{}Very strong evidence for model 5} & {\small{}0.4128} & {\small{}Negligible evidence for model 5} & {\small{}0.0257} & {\small{}Strong evidence for model 4}\tabularnewline
{\small{}$B_{36}$} & {\small{}$1.9579\times10^{-11}$} & {\small{}Very strong evidence for model 6} & {\small{}4.3528} & {\small{}Positive evidence for model 3} & {\small{}0.0853} & {\small{}Positive evidence for model 6}\tabularnewline
{\small{}$B_{45}$} & {\small{}$1.9623\times10^{-10}$} & {\small{}Very strong evidence for model 5} & {\small{}7.5112} & {\small{}Positive evidence for model 4} & {\small{}2.3862} & {\small{}Negligible evidence for model 4}\tabularnewline
{\small{}$B_{46}$} & {\small{}$1.1287\times10^{-8}$} & {\small{}Very strong evidence for model 6} & {\small{}79.2000} & {\small{}Strong evidence for model 4} & {\small{}7.9181} & {\small{}Positive evidence for model 4}\tabularnewline
{\small{}$B_{56}$} & {\small{}57.5192} & {\small{}Strong evidence for model 5} & {\small{}10.5442} & {\small{}Positive evidence for model 5} & {\small{}3.3183} & {\small{}Positive evidence for model 5}\tabularnewline
\hline 
\end{tabular}{\small\par}
\end{sidewaystable}
\newpage{}
\begin{table}[H]
\caption{Posterior model probabilities.}
\label{Flo:PMP_LAPLACE}\medskip{}

\centering{}%
\begin{tabular}{|c|>{\centering}m{2.5cm}|>{\centering}m{2.5cm}|>{\centering}m{2.5cm}|>{\centering}m{2.5cm}|>{\centering}m{2.5cm}|>{\centering}m{2.5cm}|}
\hline 
 & \multicolumn{2}{>{\centering}p{5cm}|}{\textbf{Laplace-Metropolis estimates}} & \multicolumn{2}{>{\centering}p{5cm}|}{\textbf{Harmonic mean estimates}} & \multicolumn{2}{>{\centering}m{5cm}|}{\textbf{PPD estimates}}\tabularnewline
\hline 
$ij$ & PMP for model $m_{i}$ & PMP for model $m_{j}$ & PMP for model $m_{i}$ & PMP for model $m_{j}$ & PMP for model $m_{i}$ & PMP for model $m_{j}$\tabularnewline
\hline 
\multirow{1}{*}{{\small{}$12$}} & 0.0000 & 1.0000 & 0.5358 & 0.4642 & 0.8230 & 0.1770\tabularnewline
{\small{}$13$} & 0.9999 & 0.0001 & 0.9743 & 0.0257 & 0.9847 & 0.0153\tabularnewline
{\small{}$14$} & 0.9657 & 0.0343 & 0.6758 & 0.3242 & 0.4102 & 0.5898\tabularnewline
{\small{}$15$} & 0.0000 & 1.0000 & 0.9400 & 0.0600 & 0.6240 & 0.3760\tabularnewline
{\small{}$16$} & 0.0000 & 1.0000 & 0.9940 & 0.0060 & 0.8463 & 0.1537\tabularnewline
{\small{}$23$} & 1.0000 & 0.0000 & 0.9705 & 0.0295 & 0.9328 & 0.0672\tabularnewline
{\small{}$24$} & 1.0000 & 0.0000 & 0.6436 & 0.3564 & 0.1301 & 0.8699\tabularnewline
{\small{}$25$} & 0.9356 & 0.0644 & 0.9313 & 0.0687 & 0.2630 & 0.7370\tabularnewline
{\small{}$26$} & 0.9988 & 0.0012 & 0.9931 & 0.0069 & 0.5422 & 0.4578\tabularnewline
{\small{}$34$} & 0.0017 & 0.9983 & 0.0521 & 0.9479 & 0.0107 & 0.9893\tabularnewline
{\small{}$35$} & 0.0000 & 1.0000 & 0.2922 & 0.7078 & 0.0251 & 0.9749\tabularnewline
{\small{}$36$} & 0.0000 & 1.0000 & 0.8132 & 0.1868 & 0.0786 & 0.9214\tabularnewline
{\small{}$45$} & 0.0000 & 1.0000 & 0.8825 & 0.1175 & 0.7047 & 0.2953\tabularnewline
{\small{}$46$} & 0.0000 & 1.0000 & 0.9875 & 0.0125 & 0.8879 & 0.1121\tabularnewline
{\small{}$56$} & 0.9829 & 0.0171 & 0.9134 & 0.0866 & 0.7684 & 0.2316\tabularnewline
\hline 
\end{tabular}
\end{table}

\end{document}